\documentclass[%
 reprint,
superscriptaddress,
 amsmath,amssymb,
 aps,
pra,
]{revtex4-1}

\usepackage{graphicx}% Include figure files
\usepackage{dcolumn}% Align table columns on decimal point
\usepackage{bm}% bold math
\usepackage{physics}
\usepackage{amsmath}
\newcommand{\bonnpi}{Physikalisches Institut, University of Bonn, Nussallee 12, 53115 Bonn, Germany}
\begin{document}

\title{Theoretical methods to treat a single dissipative bosonic mode coupled  globally to an interacting many body system}

\date{\today}

\begin{abstract}
  We present two approaches capable of describing the dynamics of an interacting many body system on a lattice coupled globally to a dissipative bosonic mode. Physical realizations are for example ultracold atom gases in optical lattice coupled to a photonic mode of an optical cavity or electronic gases in solids coupled to THz cavity fields.  The first approach, applicable for large dissipation strengths and any system size, is a variant of the many-body adiabatic elimination method for investigating the long time dynamics of the system. The second method extends the time-dependent matrix product techniques to capture the global coupling of the interacting particles to the bosonic mode and its open nature. It gives numerically exact results for small to intermediate system sizes. As a benchmark for our methods we perform the full quantum evolution of a Bose-Hubbard chain coupled to a cavity mode. We show that important deviations from the mean field behavior occur when considering the full atoms cavity coupling \cite{HalatiKollath2020}.

\end{abstract}
\author{Catalin-Mihai Halati}
\affiliation{\bonnpi}
\author{Ameneh Sheikhan}
\affiliation{\bonnpi}
\author{Corinna Kollath}
\affiliation{\bonnpi}
\maketitle

\section{Introduction}

The coupling of quantum matter to quantum light has been achieved in numerous experimental platforms. Examples of such realizations are ultracold atomic gases coupled to optical cavities \cite{ BaumannEsslinger2010, KlinderHemmerich2015}, electron gases in solids coupled to THz cavities \cite{ScalariFaist2012, LiuMenon2015, ZhangKono2016}, or superconducting artificial atoms coupled to on-chip cavities \cite{NiemczykGross2010, HurSchiro2016}. These systems open the exciting possibilities to study self-organization phenomena of light and matter \cite{RitschEsslinger2013, SmolkaImamoglu2014, BayerLange2017}. Novel phenomena arise from the interplay of the long range interactions and dissipative nature induced by the cavity field and the short range interactions between the atomic degrees of freedom.

In ultracold atomic systems by additionally confining the atomic gas with external optical lattice potentials an extended Bose-Hubbard model with long-range interactions has been experimentally realized \cite{KlinderHemmerichPRL2015, LandigEsslinger2016, HrubyEsslinger2018}. In addition to the superfluid and Mott insulating phases, the long range interactions also introduce charge density wave and supersolid phases. The arising phase diagram has been studied \cite{MaschlerRitsch2005, MaschlerRitsch2008, LarsonLewenstein2008, NiedenzuRitsch2010, SilverSimons2010, VidalMorigi2010, LiHofstetter2013, ElliotMekhov2016, BakhtiariThorwart2015, FlotatBatrouni2017, LinLode2018} together with the out-of-equilibrium dynamics \cite{ChiacchioNunnenkamp2018}.

\begin{figure}[hbtp]
\centering
\includegraphics[width=.45\textwidth]{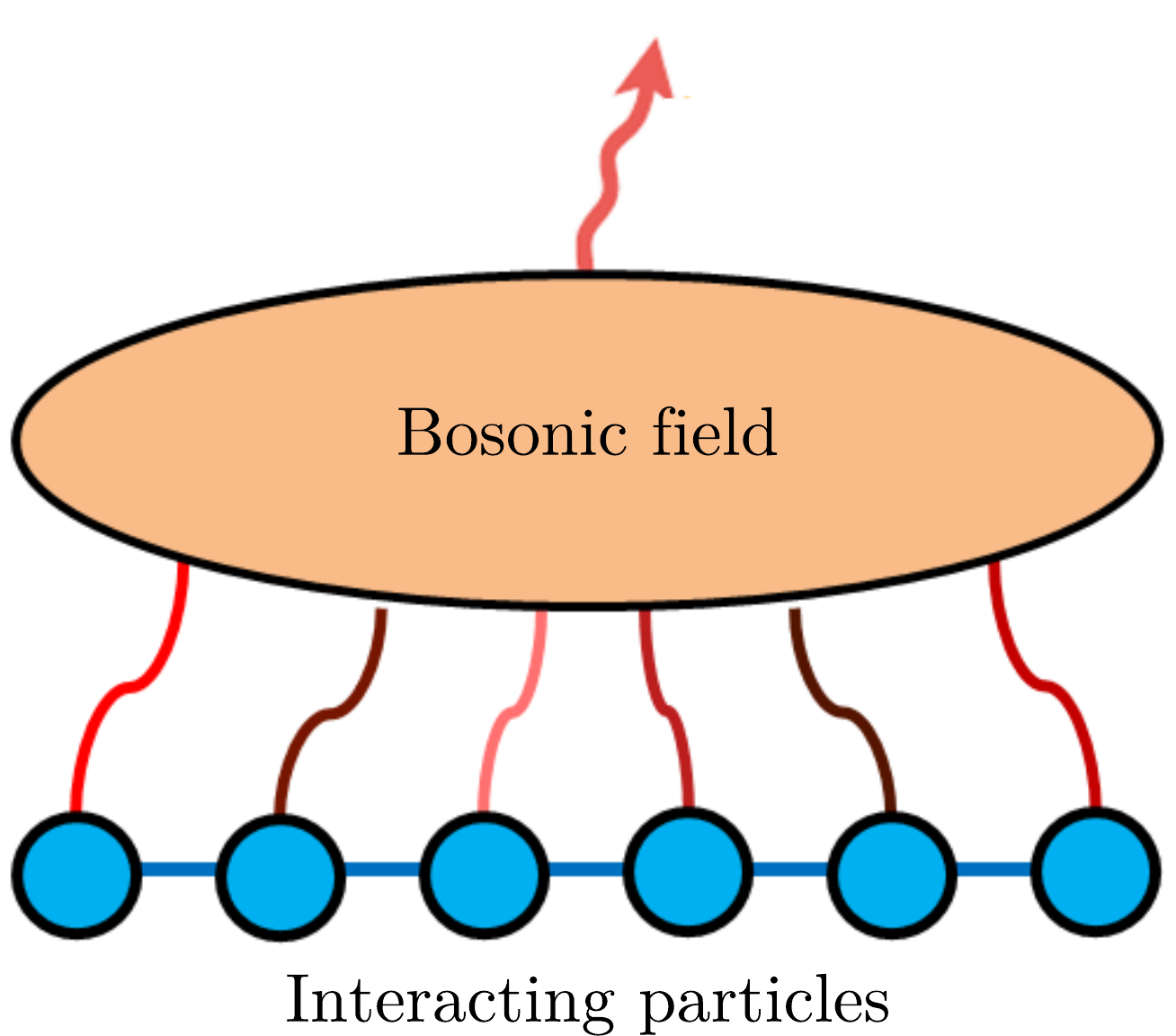}
\caption{ Sketch of a chain of interacting particles (e.g. atoms or electrons) coupled to a single bosonic quantum mode (cavity fields or phononic modes) . The bosonic mode has a dissipative nature and it is coupled to every site of the chain. The coupling strength can vary from site to site.}
\label{fig:set}
\end{figure}

The framework of most theoretical treatments of coupled atomic cavity systems so far was based on the mean field decoupling of the cavity field and the atoms \cite{RitschEsslinger2013, MaschlerRitsch2008, NagyDomokos2008}. This mean field approach simplifies the numerous technical difficulties introduced by the description of the full atom-photon coupling. Within this approach, the cavity field is assumed in a coherent state and adiabatically eliminated. This results in an effective Hamiltonian for the atoms with a self-consistency equation which is typically solved for the ground state. Deviations of this mean-field treatment have been found by taking the exact coupling between the atomic and photonic states correctly into account for small systems of one or two atoms, or two sites \cite{VukicsRitsch2007, MaschlerDomokos2007,ZhangZhou2008, KramerRitsch2014, SandnerRitsch2015, OstermannRitsch2019}, or in closed systems \cite{PiazzaZwerger2013}. This calls for new methods which can also treat larger atomic ensembles globally coupled to bosonic fields.

In this work, we develop two methods capable of capturing the exact coupling and the dissipative nature of the combined system. The first method is an extension of the many body adiabatic elimination, valid for large dissipation strengths to the combined system. Within this method any system size can be considered. It is valid for relatively long times for which system dynamics is dominated by virtual processes around the dissipation free subspace. In particular, this method provides insights about the steady state of the system.
The second method consists in quasi-exact numerical simulations based on matrix product states (MPS), which can perform efficiently the full quantum time evolution of the coupled system. This method is numerically exact and can deal with small to intermediate system sizes.

Whereas these methods are very generally applicable for quantum many body systems with short range interaction coupled globally to a single dissipative bosonic mode, we benchmark the presented methods for a Bose-Hubbard chain coupled to a cavity mode and transversely pumped with a standing-wave laser beam. We concentrate here on the description of the methods and their performance. The physical effects obtained in this system, which go beyond the established mean field results, are presented in Ref.~\cite{HalatiKollath2020}.

In Sec.~\ref{model} we describe the general setup of interacting particles on a chain coupled to a single bosonic field. Further, we describe the model for the interacting atoms coupled to an optical cavity for which the benchmarks are performed. In Sec.~\ref{mbae} we develop a variant of the many body adiabatic elimination method for the combined atom-cavity system and analyze the obtained steady state. The  numerically exact tMPS method for coupled atomic cavity systems is presented in Sec.~\ref{mps}. We discuss in detail its implementation and convergence properties.

\section{Model\label{model}}
    
In this work we consider dissipative systems of interacting particles globally coupled to a bosonic field, as sketched in Fig.~\ref{fig:set}. The particles can for example describe atoms or electrons and the bosonic quantum field can be for example a photonic field of a cavity or a long lived phononic mode. 
Generically, these systems can be described by a Lindblad equation for the density operator $\rho$ given by \cite{Carmichael1993,BreuerPetruccione2002,RitschEsslinger2013, MaschlerRitsch2008}
\begin{align}
\label{eq:Lindblad}
& \pdv{t} \rho = -\frac{i}{\hbar} \left[ H, \rho \right] + \frac{\Gamma}{2}\left(2a\rho a^\dagger-a^\dagger a \rho-\rho a^\dagger a\right).
\end{align}
where $a$ and $a^\dagger$ are the annihilation and creation operators for the bosonic mode. The dissipative term proportional to the dissipation strength $\Gamma$ takes into account the losses from the bosonic mode. It is described by a Lindblad form of the dissipator where the jump operator is the annihilation operator $a$ of the bosonic mode. In a cavity these can be due to the imperfections of the mirrors and for phononic modes it describes the decay into a bath of phononic modes. Let us note, that a generalization of the developed methods to any set of jump operators acting only on either the bosonic or the particle part of the system is straightforward.

\begin{figure}[hbtp]
\centering
\includegraphics[width=.45\textwidth]{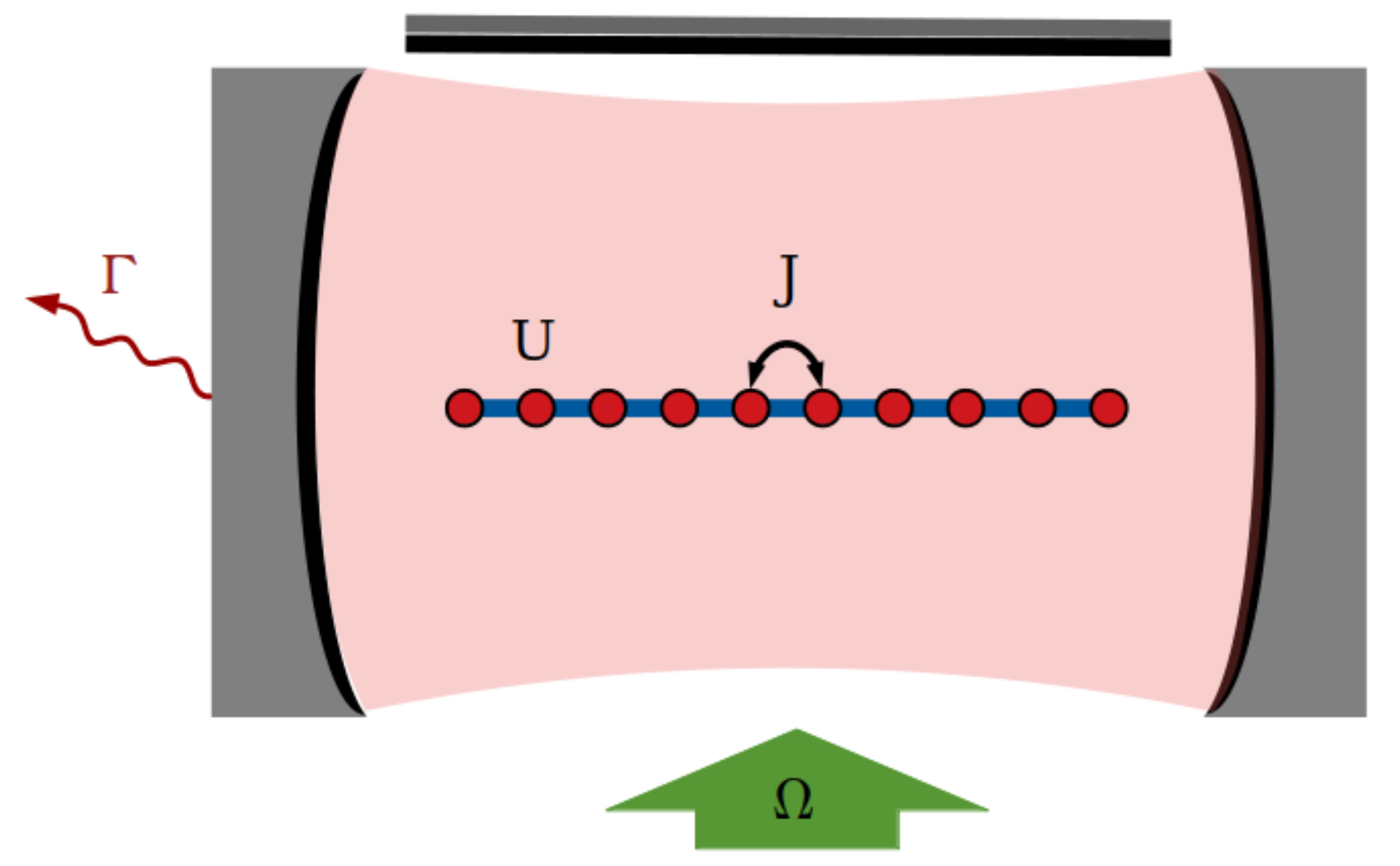}
\caption{ Sketch of the bosonic atoms confined in a one-dimensional chain in an optical cavity. The atoms tunnel with the amplitude $J$ and have an on-site interaction of strength $U$. 
The coupling of the atoms to the cavity is realized with a retroreflected transverse pump beam. As the lattice spacing is commensurate with half of the wavelength of the cavity mode, the cavity field is coupled to the total imbalance between the odd and even sites of the chain. The strength of the coupling is controlled by the pump amplitude $\Omega$. The cavity is losing photons with the dissipation strength $\Gamma$, due to the imperfections of the mirrors.
 }
\label{fig:setup}
\end{figure}

The methods that we present in this work can deal with a Hamiltonian of the following form, $H=H_c+H_{\text{chain}}+H_{\text{ac}}$. Where $H_c$ contains only operators of the bosonic field, $H_{\text{chain}}$ is an interacting short-range one dimensional Hamiltonian for the many body degrees of freedom. $H_\text{ac}\propto ( a + a^\dagger) \mathcal{A}$ couples the bosonic field to a global particle operator, where $\mathcal{A}$ is a sum over operators which act on one, or at most two, atomic sites.

We will benchmark the developed methods using interacting bosons confined to a chain coupled to a single cavity mode transversely pumped with a standing-wave laser beam, as depicted in Fig.~\ref{fig:setup}. However, the methods are easily adaptable to interacting spins or interacting fermions. In the considered model, the Hamiltonian has the form  \cite{RitschEsslinger2013, MaschlerRitsch2008, NagyDomokos2008}
\begin{align} 
\label{eq:Hamiltonian}
&H=H_\text{c}+H_{\text{atom}}+H_{\text{ac}} \\
&H_\text{c}= \hbar\delta a^\dagger a,\nonumber\\
&H_\text{atom}=H_{\text{int}}+H_{\text{kin}},\nonumber\\
&H_{\text{int}}=\frac{U}{2} \sum_{j=1}^L n_{j}(n_{j}-1),\nonumber\\
&H_{\text{kin}}=-J \sum_{j=1}^{L-1} (b_{j}^\dagger b_{j+1} + b_{j+1}^\dagger b_{j}), \nonumber\\
&H_{\text{ac}}=  -\hbar\Omega ( a + a^\dagger) \Delta, ~~ \Delta=\sum_{j=1}^L (-1)^j n_j \nonumber.
\end{align}
The term $H_\text{c}$ describes the cavity mode in the rotating frame of the pump beam, with a detuning between the cavity mode and the transverse pump beam $\delta=\omega_\text{c}-\omega_\text{p}$.
The operators $b_{j}$ and $b_{j}^\dagger$ are the bosonic annihilation and creation operators of the atoms on site $j$ and  $n_{j}=b_{j}^\dagger b_{j}$. $L$ denotes the number of sites of the chain and the total number of bosonic atoms is $N$. For the atomic part of the Hamiltonian we have the terms $H_{\text{kin}}$ which describes the tunneling processes of the atoms  with the amplitude $J$ and the term $H_{\text{int}}$ representing the repulsive on-site interaction of strength $U>0$. 
The term $H_\text{ac}$ gives the coupling between the cavity field and the total imbalance between the odd and even sites of the chain, $\Delta$, with the effective pump amplitude $\Omega$. This coupling is realized due to the assumed commensurability of the cavity mode with twice the periodicity of the lattice spacing within the chain \cite{MaschlerRitsch2008}.  
The Hamiltonian, Eq.~(\ref{eq:Hamiltonian}), exhibits a $\mathbb{Z}_2$ symmetry associated with the inversion of the sign of the cavity field, $a$, and the atomic odd-even imbalance, $\Delta$. 
However,  the $\mathbb{Z}_2$ symmetry is only a weak symmetry of the Liouvillian \cite{BucaProsen2012,AlbertJiang2016}, since the transformation does not commute with the jump operator $a$ of the Lindblad equation, Eq.~(\ref{eq:Lindblad}). Thus, a zero expectation value for the cavity field is expected in the steady state of the system \cite{VukicsRitsch2007, MaschlerDomokos2007, SandnerRitsch2015, GammelmarkMolmer2012}. 

The methods and results presented in this work are to be put in contrast with the approach of adiabatically eliminating the cavity field by a mean field decoupling of the atoms and the cavity mode \cite{RitschEsslinger2013}. 
In this crude approximation, after eliminating the cavity field one obtains an effective Hamiltonian for the atoms, which for the system presented in Eqs.~(\ref{eq:Lindblad})-(\ref{eq:Hamiltonian}), is given by $H_{\text{eff}}=H_{\text{kin}}+H_{\text{int}}-V_\text{c} \Delta$. 
The parameter $V_\text{c}$ has to be determined self-consistently as it depends on the expectation value of odd-even imbalance, $V_\text{c}=\frac{2\hbar\Omega^2\delta}{\delta^2 +\Gamma^2/4}\langle \Delta \rangle$. 
In this approach above a certain threshold $\Omega_{\textrm{MF},c}\sqrt{N}$ the cavity field $\langle a \rangle$ takes a finite value and the atoms self-organize into a density modulated pattern either on the odd or even sites of the chain breaking spontaneously the $\mathbb{Z}_2$ symmetry of the effective Hamiltonian. The steady state is a pure state composed of a product state between the atomic and photonic sector $\rho_{\textrm{MF}}=\ket{\alpha(\Delta_{\text{eff}}),\Delta_{\text{eff}}}\bra{\alpha(\Delta_{\text{eff}}),\Delta_{\text{eff}}}$. The photonic mode is in the coherent state $\alpha(\Delta_{\text{eff}})$ with $\alpha(\Delta)=\frac{\Omega}{\delta-i \Gamma/2}\Delta$ and the corresponding average photon number is $n_{\textrm{MF}}=\frac{\Omega^2}{\delta^2+\Gamma^2/4}\Delta_{\text{eff}}^2$. The atomic state $\ket{\Delta_\text{eff}}$ denotes the ground state of the effective Hamiltonian with the self-consistency condition. The effective imbalance $\Delta_{\text{eff}}$ is defined as the expectation value of the odd-even imbalance in the ground state of the effective Hamiltonian.

\section{Many body adiabatic elimination formalism\label{mbae}}
In order to understand the long-time behavior of our system in the strongly dissipative regime, we employ the many body adiabatic elimination method \cite{RipollCirac2009, PolettiKollath2013, ReiterSorensen2012, Kessler2012, SciollaKollath2015}. In this section we describe the many body adiabatic elimination formalism and how to apply it to the photon mode coupled to the interacting atoms. 

\begin{figure}[hbtp]
\centering
\includegraphics[width=.4\textwidth]{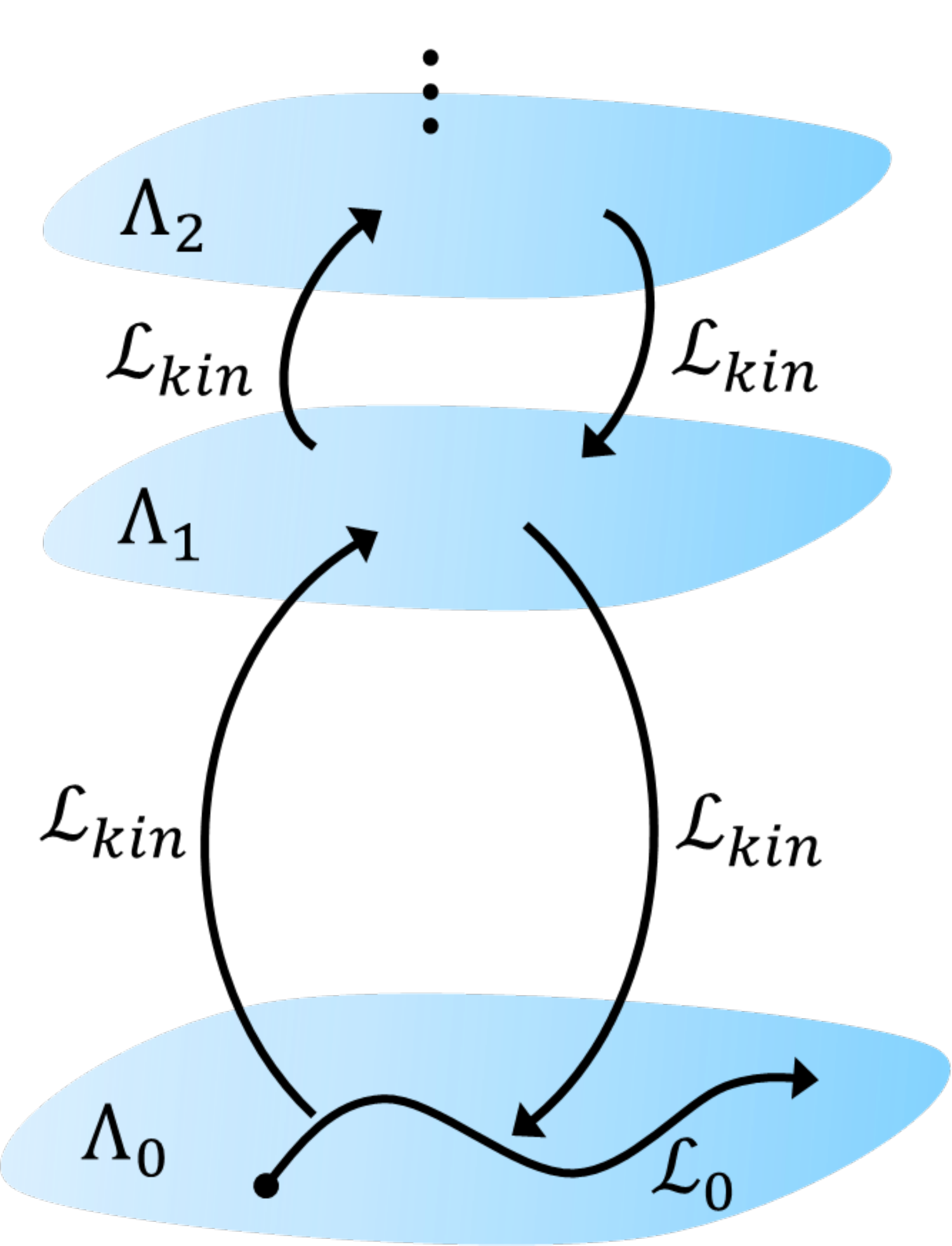}
\caption{Sketch of the spectrum of the Liouvillian $\mathcal{L}_0$. The subspaces $\Lambda_\alpha$ are spanned by the eigenstates of $\mathcal{L}_0$ which have eigenvalues with the same real part. The subspace $\Lambda_0$ is the decoherence free subspace of $\mathcal{L}_0$, containing only states with a vanishing real part of the eigenvalues. While the evolution given by $\mathcal{L}_0$ is contained within a subspace, the Liovillian $\mathcal{L}_\text{kin}=-\frac{i}{\hbar}[H_{\text{kin}},\cdot]$ can induce transitions between the different subspaces $\Lambda_\alpha$.
 }
\label{fig:dfs}
\end{figure}

\subsection{\label{sec:ss}Derivation of the equation of motion}

We assume that we can consider the kinetic energy term, $H_{\text{kin}}$, as a perturbation  ($\hbar\Gamma\gg \hbar\Omega, \hbar\delta \gg J$) compared to the other terms in the Liouvillian $\mathcal{L}_0=-\frac{i}{\hbar}[H_\text{c}+H_{\text{int}}+H_{\text{ac}},\cdot]+\mathcal{D}(\cdot)$.
This approach will give an insight into the effective dynamics of the density matrix in the decoherence free subspace of $\mathcal{L}_0$, i.e.~the space $\Lambda_0$ formed by all density matrices $\rho^0$ which are eigenstates of the superoperator $\mathcal{L}_0$ with vanishing real part of the eigenvalue. The other spaces, $\Lambda_{m}$, formed by the right eigenvectors  corresponding to eigenvalues with equal non-zero real part are only considered within perturbation theory.

In Fig.~\ref{fig:dfs} we sketch the decoherence free subspace $\Lambda_0$ and two different subspaces $\Lambda_1$, $\Lambda_2$, together with the action of the Liovillian $\mathcal{L}_0$ and the perturbation $\mathcal{L}_\text{kin}=-\frac{i}{\hbar}[H_{\text{kin}},\cdot]$ which connects the different subspaces.
If we consider only contributions from the subspace, $\Lambda_1$, that can be accessed via one hopping event, the effective dynamics for the elements of the decoherence free subspace  is given by \cite{PolettiKollath2013, SciollaKollath2015}
\begin{align}
\label{eq:decfree0}
\pdv{t} \rho^{0} \approx \lambda_0 \rho^0+\frac{1}{\hbar^2} P_0 \left[ H_{\text{kin}},\mathcal{L}_0^{-1} P_1 \left[H_{\text{kin}},\rho^0\right]\right],
\end{align}
where $\rho^0 \in \Lambda_0$ is an eigenstate of $\mathcal{L}_0$ with vanishing real part of the eigenvalue, i.e.~$\mathcal{L}_0[\rho^0]=\lambda_0\rho^0$ with $\Re(\lambda_0)=0$.  The operators $P_0$ and $P_1$ are the projectors onto the subspaces $\Lambda_0$ and $\Lambda_1$, respectively.

In the following we need to determine the elements of the decoherence free subspace $\Lambda_0$ and of $\Lambda_1$. Solving the eigenvalue equation belonging to $\mathcal{L}_0$ is already complex for the system we consider. However, we find that a set of right eigenstates of $\mathcal{L}_0$ is given by states of the form 
\begin{align}
\label{eq:ansatz_u}
& \rho=\ket{\alpha(\Delta);n_1,\dotso,n_L}\bra{\alpha(\Delta');n_1',\dotso,n_L'}.
\end{align}
At this point we do not assure that these states are physical density matrices. 
The atomic part is given by Fock states with the odd-even imbalances $\Delta=\sum_{j} (-1)^j n_j$ and $\Delta'=\sum_{j} (-1)^j n_j'$ and its total interaction energies $u= \frac{U}{2}\sum_{j}n_j(n_j-1)$ and $u'= \frac{U}{2}\sum_{j}n_j'(n_j'-1)$.
The photons are in a coherent state which depends on the atomic imbalance
\begin{align}
\label{eq:coherent}
 \alpha(\Delta) &=\frac{\Omega}{\delta-i\Gamma/2} \Delta.
\end{align}

The corresponding eigenvalues for the right eigenvectors in Eq.~(\ref{eq:ansatz_u}) are given by 
\begin{align}
\label{eq:eigenvalue_u}
\lambda(\Delta,u,\Delta',u')& = - \frac{1}{2}\frac{\Omega^2\Gamma}{\delta^2+\Gamma^2/4}(\Delta-\Delta')^2 \\
&+i\left[\frac{\Omega^2\delta}{\delta^2+\Gamma^2/4}(\Delta^2-\Delta'^2)-(u-u')\right]. \nonumber
\end{align}
For $\Delta=\Delta'$ the real part of the eigenvalues is zero. Thus,  the states in Eq.~(\ref{eq:ansatz_u}) with $\Delta=\Delta'$ lie in the decoherence free subspace of $\mathcal{L}_0$. Interestingly, the eigenstates with $\Delta=\Delta'$, but with different interaction energies $u\neq u'$ have purely imaginary eigenvalues.
The subspace which can be accessed via a hopping event from the decoherence free subspace is given for the states in which $\Delta=\Delta'\pm 2$. 

By writing explicitly the equations of motion, Eq.~(\ref{eq:decfree0}), for the elements of the decoherence free subspace (see Appendix A) one obtains that the mixed state given by
\begin{align}
\label{eq:ss_gen}
\rho_{\text{mix}} &=\frac{1}{\mathcal{N}}\sum_{\{n_j\}} \ket{\alpha(\Delta);n_1,\dotso,n_L}\bra{\alpha(\Delta);n_1,\dotso,n_L}
\end{align}
is a steady state of the system. Here we sum over all possible density configurations $\{n_j\}$ and $\mathcal{N}$ is the number of these configurations, which is the number of ways one can arrange $N$ identical particles in $L$ sites, $\mathcal{N}=\begin{pmatrix} L+N-1 \\ N \end{pmatrix}$. 

\begin{figure}[hbtp]
\centering
\includegraphics[width=.5\textwidth]{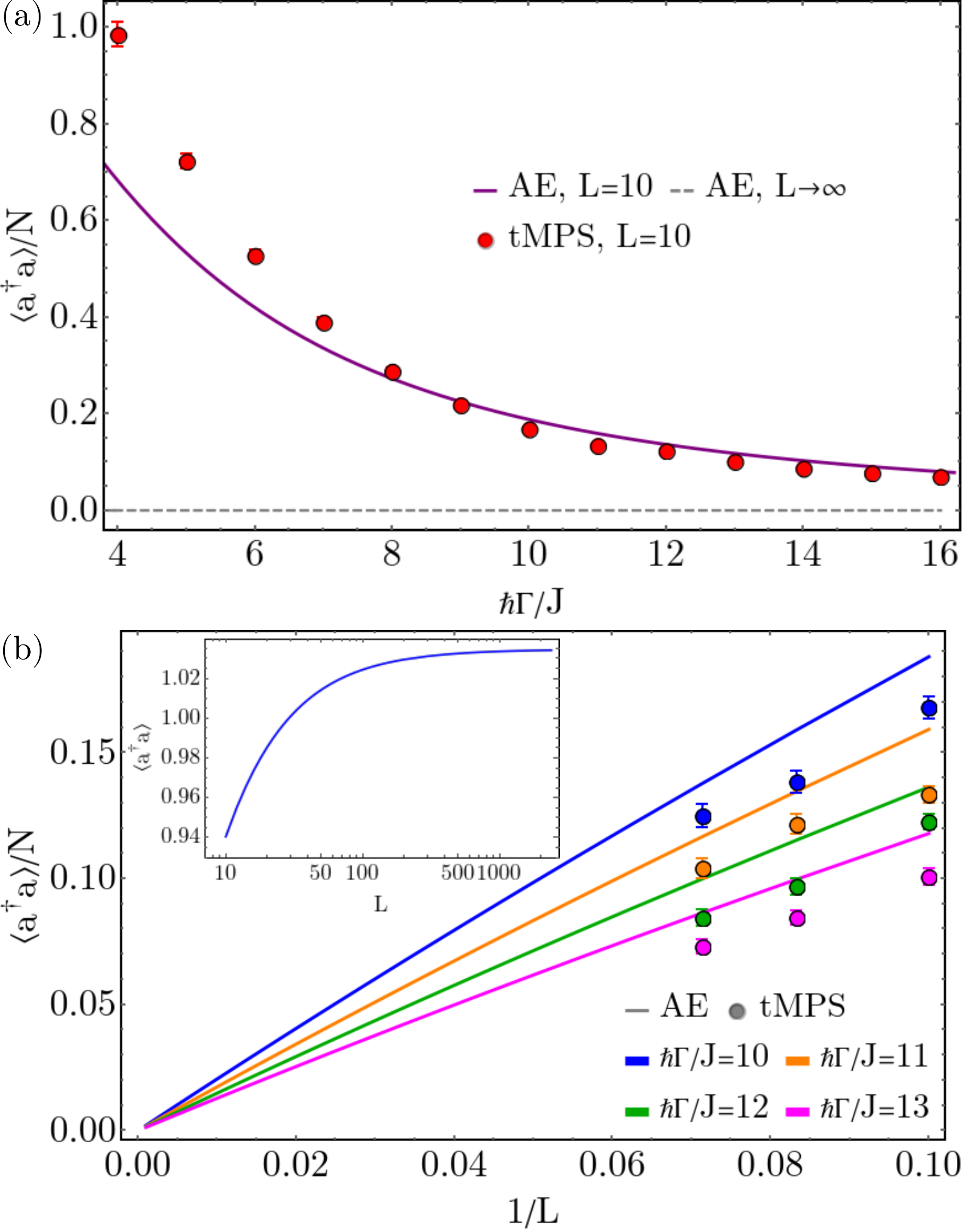}
\caption{
(a) The dependence of the scaled photon number $\langle a^\dagger a\rangle/N$, on the dissipation strength, $\hbar \Gamma/J$, using tMPS and many-body adiabatic elimination, for $\hbar\delta/J=2$, $U/J=2$ and $\hbar\Omega\sqrt{N}/J=4.47$.
(b) The dependence of the scaled photon number $\langle a^\dagger a\rangle/N$, on $1/L$.
The parameters used are $N=L/2$ particles and $\hbar\Gamma/J\in\{10,11,12,13\}$. The behavior is consistent with a $L^{-1}$ scaling of $\langle a^\dagger a\rangle/N$.
In the inset we have the dependence of the photon number, $\langle a^\dagger a\rangle$, on the system size, $L$, which seems to saturate at large $L$.
 }
\label{fig:ae}
\end{figure}

In order to illustrate the range of validity of the many-body adiabatic elimination approach, we compare its results with the numerically exact tMPS method results, which will be introduced and discussed in Sec.~\ref{mps}. In Fig.~\ref{fig:ae}(a) we observe that the scaled photon number computed with $\rho_{\text{mix}}$ is in agreement with the tMPS results for $\hbar\Gamma/J\gtrsim 9$, which is larger than the other chosen parameters, and the agreement continues as we increase the dissipation strength.

In Ref.~\cite{ChiacchioNunnenkamp2018} the authors consider the same model, but  they eliminate the cavity field and analyze the obtained effective Liouvillian in the atomic sector. As in their case the effective jump operators are Hermitian, it follows directly that the fully mixed state is a steady state. In contrast, in our analysis we consider the full Liouvillian Eqs.~(\ref{eq:Lindblad}) and (\ref{eq:Hamiltonian}), including the photonic degrees of freedom, and because the jump operator (annihilation operator of the cavity mode $a$) is not Hermitian we need to perform the complicated many body adiabatic elimination in order to obtain insights into the nature of the steady state. 

\subsection{\label{sec:thermo}Properties of the steady state}

With the many body adiabatic elimination method we obtain a steady state $\rho_{\text{mix}}$ which is very different in nature compared with the expected mean field state. The mean field state is the ground state of the effective Hamiltonian and by this a pure state with a coherent state in the photonic sector and a density wave in the atomic sector. In contrast, tracing out the photonic mode in $\rho_{\text{mix}}$ leads to a fully mixed atomic sector corresponding to an infinite temperature state, which is very different from a pure ground state.
Moreover, the steady state, $\rho_{\text{mix}}$, is a mixture of separable states, thus no entanglement is present between the photons and the atoms, but the strong cavity-atoms coupling is reflected by the fact that in each of the pure states present in the mixture the cavity field is fully determined by the atomic density profile.

This very distinct nature of the steady state is also reflected in the physical observables. Due to the fully mixed atomic sector, the density-density correlations have a flat profile. Therefore, the staggering of the density-density correlations vanishes. $\rho_{\text{mix}}$ has a zero expectation value of the cavity field, $\langle a\rangle=0$, capturing the weak $\mathbb{Z}_2$ symmetry of the system \cite{HalatiKollath2020}, but has a finite expectation value of the photon number. The average photon number for the state $\rho_{\text{mix}}$, Eq.~(\ref{eq:ss_gen}), is given by
\begin{align}
\label{eq:ae_pho}
\langle a^\dagger a\rangle=\sum_{\Delta} \frac{(\Omega \sqrt{N})^2}{\delta^2+\Gamma^2/4} \frac{c_{\Delta}}{\mathcal{N}} \frac{\Delta^2}{N},
\end{align}
where the sum is taken over the set $\Delta\in\{-N,-N+2,\ldots,N-2,N\}$ and $c_\Delta$ being the number of states with a certain imbalance $\Delta$, given by 
\begin{align} 
\label{eq:ae_state_no}
c_\Delta=\begin{pmatrix} \frac{1}{2}(L+N+\Delta)-1 \\  \frac{1}{2}(N+\Delta) \end{pmatrix} \begin{pmatrix} \frac{1}{2}(L+N-\Delta)-1 \\  \frac{1}{2}(N-\Delta) \end{pmatrix}.
\end{align} 

By plotting the scaled photon number $\langle a^\dagger a\rangle/N$ at a fixed filling $N/L$, Fig.~\ref{fig:ae}(b), we can see that this quantity vanishes as $L^{-1}$ at large $L$. This implies that even though the scaled photon density per atom is finite for any finite size system, it goes to zero in the thermodynamic limit, $L\to\infty$. Thus, the many body adiabatic elimination method tells us that in the thermodynamic limit at large dissipation strengths the system is no longer in a superradiant state with a finite $\langle a^\dagger a\rangle/N$, but in a state with an average number of zero photons and a fully mixed atomic sector. This is very distinct to the normal state predicted by the mean-field approach. Despite the fact that in both states the average photon number vanishes, the atomic part of  the mean-field state is a pure state and not the infinite temperature state as predicted by adiabatic elimination. Therefore, our results also question the nature of the transition predicted by the mean-field approach between the superradiant state and the normal state.

\section{Time-dependent matrix product state (\MakeLowercase{t}MPS) method for combined atom-cavity systems\label{mps}}

In this section we describe the numerical exact method based on matrix product states (MPS) we developed to perform the quantum time evolution of the combined system. 

\subsection{\label{mps2}Details of the tMPS method for the coupled photon-atom system}

The considered dissipative system of atoms coupled to an optical cavity poses several challenges for its treatment via MPS based methods. The first difficulty arises due to the in principle arbitrarily large dimension of the Hilbert space of the cavity field. The second obstacle is the global coupling of the cavity mode to the interacting atoms. The third challenge is the dissipative nature of the combined system due to the photon losses. In the following we describe how our implementation overcomes all these difficulties.
We implement the newly developed algorithm efficiently using the ITensor library \cite{itensor, FishmanStoudenmire2020}.

We begin by presenting how the dissipative aspect of the considered models is included in the numerical method. For the simulation of the dissipative many body quantum systems the time-evolution of the density matrix following the Lindblad equation needs to be determined. State of the art are two different routes: the first is the purification approach which relies on the rewriting of the density matrix with a larger dimension \cite{VerstraeteCirac2004, ZwolakVidal2004}. The second is the stochastic unravelling of the master equation using quantum trajectories \cite{DalibardMolmer1992, GardinerZoller1992}. This approach has the advantage of simulating the time-evolution of wavefunctions instead of density matrices at the disadvantage of a stochastic sampling.
We have chosen as a first implementation the stochastic unraveling of the master equation. This has in particular two reasons: First, already the representation of the interacting ground state of the bosonic atoms as initial state would have been demanding in the purification approach. Secondly, the additional presence of the large Hilbert space of the photons would have increased the required matrix dimension.

In the stochastic formulation we take good quantum numbers into account for the atomic sector. The efficient combination of the stochastic unraveling of the master equation with the matrix product state methods has been relatively recent and only few groups have efficient implementations taking conserved quantum numbers into account (see e.g.~\cite{ Daley2014, BernierKollath2013, BonnesLauchli2014, WallRey2016, ManzoniDouglas2017, JaschkeCarr2018, TindallJaksch2019}).

\begin{figure}[hbtp]
\centering
\includegraphics[width=.38\textwidth]{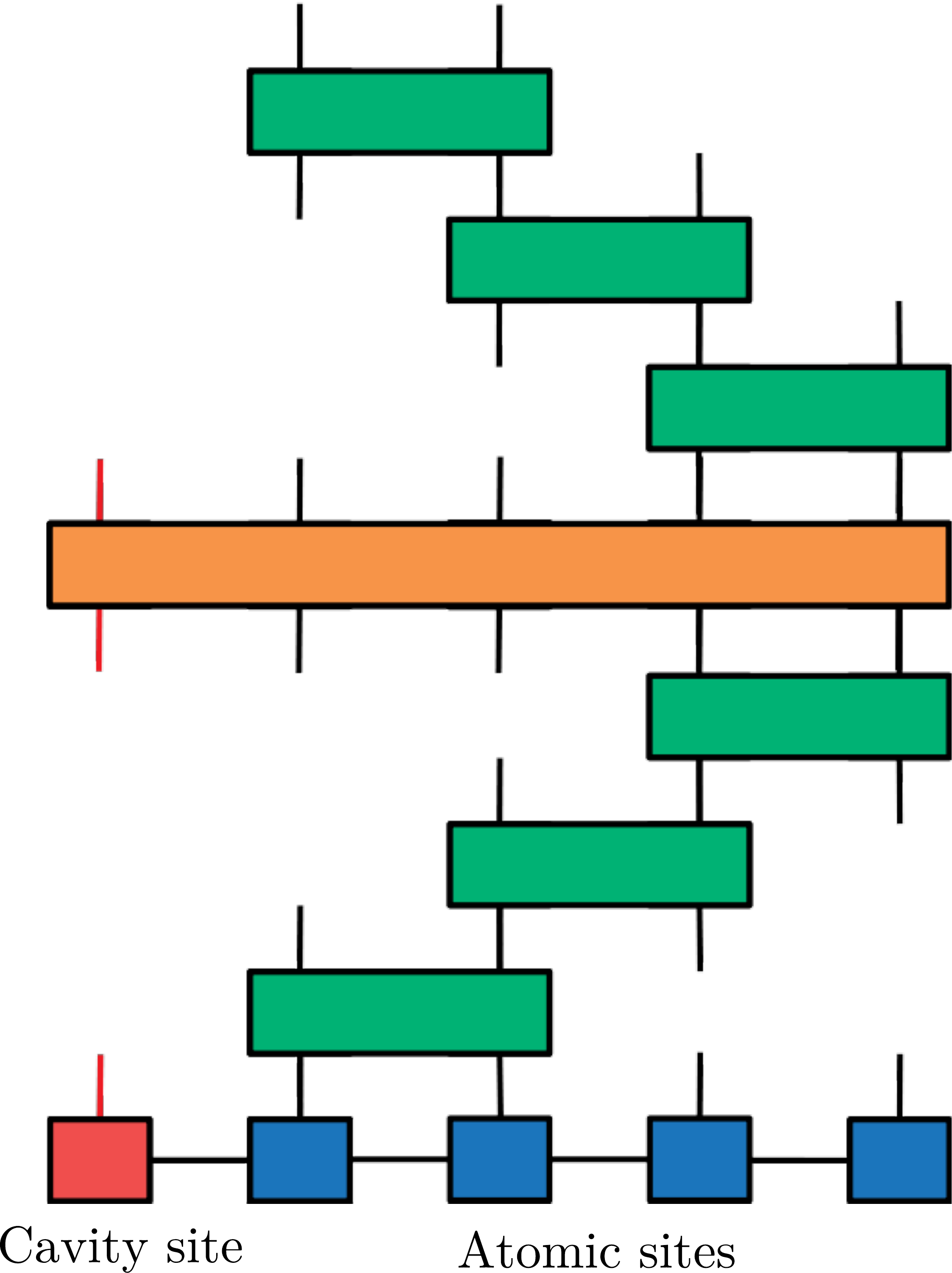}
\caption{
The graphical representation of one step of the evolution in time, based on the Trotter-Suzuki decomposition described in the text, Eq.~\ref{eq:trott1}. The first (red) site in the graphical representation of the MPS structure corresponds to the cavity mode and the rest to the atomic sites. To be noted that the cavity mode index marked with a red line has a large local dimension. Green boxes represent the application of the two site gates of the atomic terms of the time-evolution after Trotter-Suzuki decomposition followed by an SVD compression step. With orange box we depict the large tensor corresponding to the time evolution of the cavity and cavity-atoms coupling terms of the Hamiltonian. Its application is detailed in Fig.~\ref{fig:mps4}.
 }
\label{fig:mps2}
\end{figure}

In order to apply the stochastic unravelling procedure, the time-evolution of many trajectories of pure states is sampled and finally the results are averaged. The initial states for the trajectories are drawn corresponding to their probability weights in the initial density matrix. Then a stochastic time-evolution is performed for each trajectory which is described in the following:
\begin{itemize}
\item A random number $\eta$ is drawn from the interval $\left[0,1\right)$.
\item For each trajectory, the time evolution is performed for a time step with a non-unitary time evolution operator, corresponding to the effective Hamiltonian, $\tilde{H}=H-\frac{i}{2}\hbar\Gamma a^\dagger a$. 
\item  Since the effective Hamiltonian is not Hermitian, this leads to a decay of the norm of state in time. The non-unitary deterministic time evolution is performed until the norm is smaller than a threshold posed by the random number $\eta$. 
\item A quantum jump is performed by applying the jump operator $a$ onto the wavefunction and the state is normalized. 
\item The described procedure is repeated until the required final time is reached. 
\end{itemize}

One can show \cite{DalibardMolmer1992, GardinerZoller1992} that taking the Monte Carlo average over all sampled quantum trajectories, the described time evolution reproduces the Lindblad dynamics correctly up to the first order in the chosen time step. In order to achieve convergence in the computed quantities many trajectories are required and the time steps need to be chosen small enough to avoid multiple jumps within one time step. In our case, because the jump operator only acts on the photonic space, we need to sample several hundred trajectories, as discussed in the Sec.~\ref{sec:conv}.

\begin{figure}[hbtp]
\centering
\includegraphics[width=.48\textwidth]{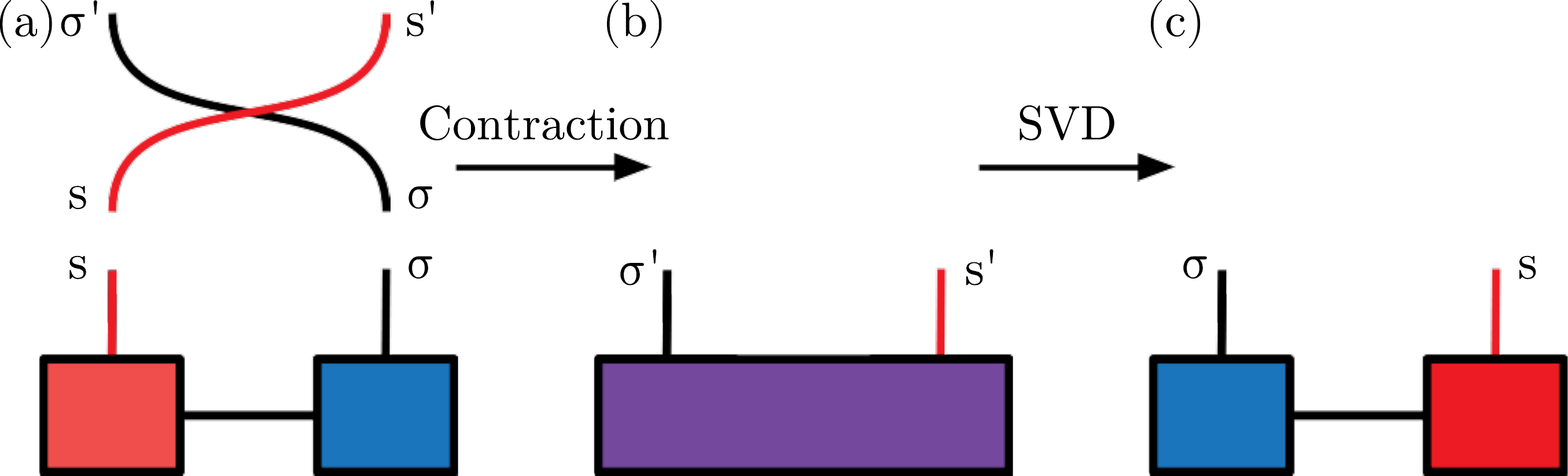}
\caption{
The graphical representation of the application of the swap gate procedure: (a) The two site MPS with the physical indices $s$, corresponding to the cavity site, and $\sigma$, corresponding to the atomic site, and the swap gate with the indices $\left(s,\sigma,\sigma',s'\right)$. (b) The application of the swap gate onto the MPS by contracting the indices $s$, $\sigma$ and the MPS bond index. (c) Restoring the MPS structure by performing a SVD and renaming the indices $\sigma'\to\sigma$ and $s'\to s$.
 }
\label{fig:mps3}
\end{figure}

In order to perform the time evolution within the MPS formalism, we represent the wave function as a matrix product state (MPS) \cite{Schollwock2011}, with the first site initially corresponding to the cavity mode and the rest to the atomic lattice using a Fock basis for each site (see Fig.~\ref{fig:mps2}). In section \ref{sec:center}, we compare this geometry to the one, where the cavity site is placed in the center of the chain. In order to take care of the in principle arbitrarily large Hilbert space of the photonic mode, we introduce a cutoff for the dimension of the local Hilbert space of the photonic site, 
which is dynamically adapted during the time evolution. This is done by setting a truncation goal of the photonic distribution and the details are given in Sec.~\ref{sec:photon}. For benchmarking we also present results in which a fixed dimension of the photonic Hilbert space is used. 

The global range coupling between the cavity mode and all the atomic sites makes the use of the tMPS implementation for short-range Hamiltonians  based on the Trotter-Suzuki decomposition impossible. Thus, in order to take both the global coupling between photons and atoms and the short range interaction of the atoms into account, we develop a variant of the tMPS based on the dynamical deformation of the MPS structure. The dynamical deformation allows one to alter the order of the sites as needed using swap gates \cite{StoudenmireWhite2010, Schollwock2011, WallRey2016}. Previous variants of the MPS time evolution with swap gates dealt with short-range interaction in two dimensional models \cite{StoudenmireWhite2010}, or spin-boson models \cite{WallRey2016, WallRey2017}. Our implementation, in contrast, can efficiently deal with interacting bosonic models globally coupled to the dissipative photonic field. An adaptation to fermionic and spin systems coupled to photonic modes is straightforward.

\begin{figure}[hbtp]
\centering
\includegraphics[width=.3\textwidth]{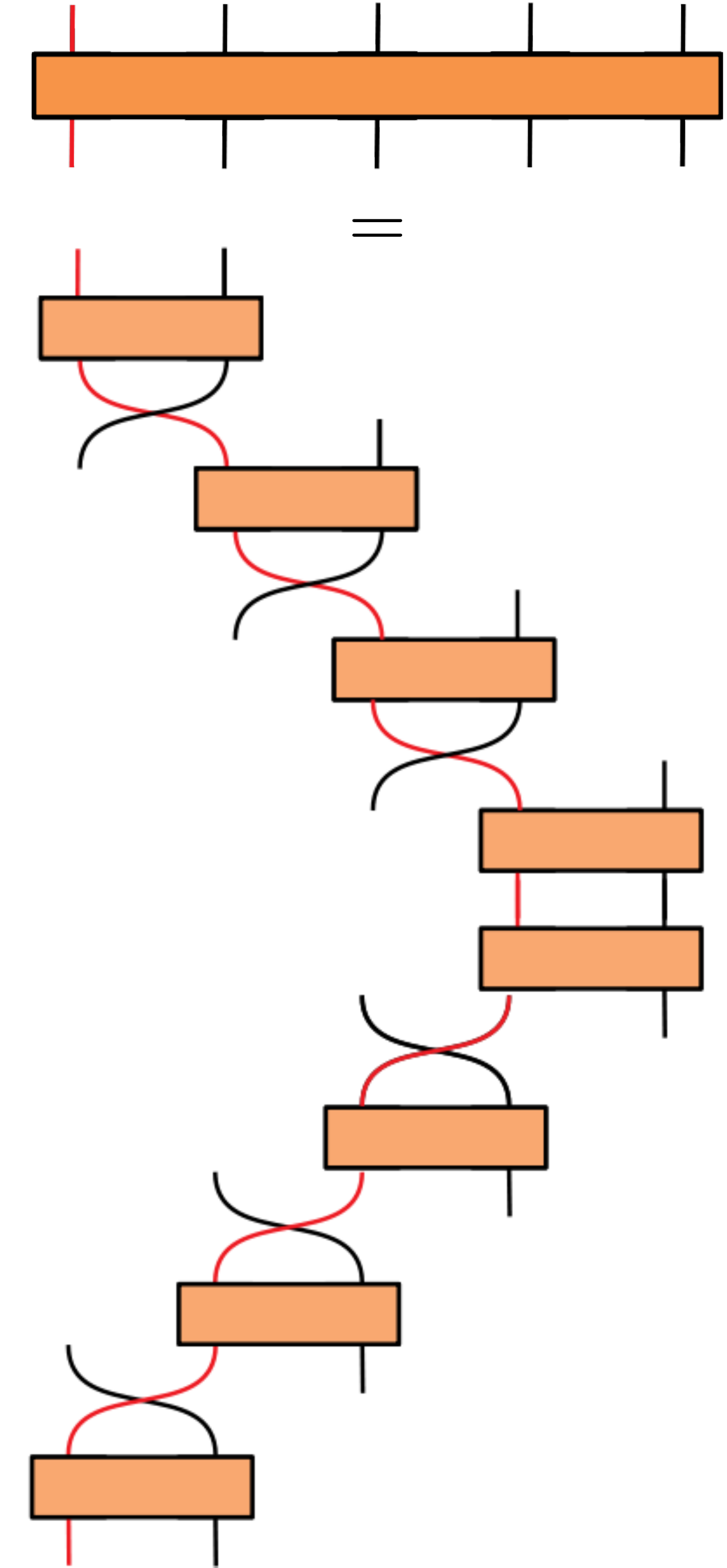}
\caption{
The graphical representation of application of the Trotter-Suzuki decomposition of the terms containing the cavity field, Eq.~\ref{eq:trott2}. Swap gates are needed to bring the initially distant sites close to each other. 
 }
\label{fig:mps4}
\end{figure}

In the following, we describe our procedure for performing a time step $\textrm{d}t$ with the effective Hamiltonian, $\tilde{H}$. It is based on the Trotter-Suzuki decomposition of the time evolution propagator in combination with swap gates. The terms are split in order to separate the terms containing the cavity field operators and the remaining terms
\begin{align}
\label{eq:trott1}
&e^{-\frac{i\mathrm{d}t}{\hbar} \tilde{H}}\approx \\
&~ e^{-\frac{i\mathrm{d}t}{2\hbar} \left(H_{\text{kin}}+H_{\text{int}}\right)}e^{-\frac{i\mathrm{d}t}{\hbar} \left(H_{\text{ac}}+H_\text{c}-\frac{i}{2}\hbar\Gamma a^\dagger a\right)}e^{-\frac{i\mathrm{d}t}{2\hbar} \left(H_{\text{kin}}+H_{\text{int}}\right)}. \nonumber
\end{align}
This decomposition is valid to the order $\mathcal{O}(\mathrm{d}t^3)$ in the time-step. The evolution given by the operator $e^{-\frac{i\mathrm{d}t}{2\hbar} \left(H_{\text{kin}}+H_{\text{int}}\right)}$ which only contains the atomic operators is computed as in the standard tMPS algorithm for short-range interactions \cite{DaleyVidal2004, WhiteFeiguin2004} by a further decomposition into two site gates. The two site gates are applied to the MPS followed by a compression step via a singular value decomposition (SVD) in the order sketched in Fig.~\ref{fig:mps2}.
We mention that the boundary gates at the beginning and the end of the bosonic chain differ from the gates applied in the bulk.

For the operator $e^{-\frac{i\mathrm{d}t}{\hbar} \left(H_{\text{ac}}+H_{\text{c}}-\frac{i}{2}\hbar\Gamma a^\dagger a\right)}$  which contains the global coupling to the cavity field, we use the fact that we can decompose $H_{\text{ac}}$ such that each term only acts on two sites --even though distant ones--

\begin{align}
\label{eq:trott2}
&e^{-\frac{i\mathrm{d}t}{\hbar} \left(H_{\text{ac}}+H_{\text{c}}-\frac{i}{2}\hbar\Gamma a^\dagger a\right)}= \\
&\quad=\prod_{j=L}^{1} e^{-\frac{i\mathrm{d}t}{2} \left[-\Omega (a+a^\dagger)(-1)^{j}n_j+\frac{1}{L}\left(\delta-\frac{i}{2}\Gamma\right)  a^\dagger a\right]} \times \nonumber \\
&\quad\quad\prod_{j=1}^{L} e^{-\frac{i\mathrm{d}t}{2} \left[-\Omega (a+a^\dagger)(-1)^{j}n_j+\frac{1}{L}\left(\delta-\frac{i}{2}\Gamma\right)  a^\dagger a\right]} +\mathcal{O}(L\mathrm{d}t^3). \nonumber 
\end{align}
This means that we need to apply two-site operators where the two sites are not neighbors in the initial MPS representation. In order to solve this problem, we adapt the structure of the MPS while applying the time-evolution gates such that we bring the two sites on which the operator acts next to each other. This approach is implemented using swap gates, where the action of the swap gates consists in the swapping of the physical indices of two neighboring MPS matrices, i.e.
\begin{align}
\label{eq:swap}
&S_{s,\sigma_i}\left(M^{\sigma_1}...M^{s}M^{\sigma_{i}}...M^{\sigma_L}\right)  \\
&=M^{\sigma_1}...(MM)^{\sigma_{i},s}...M^{\sigma_L} \nonumber \\
&=M^{\sigma_1}...M^{\sigma_{i}}M^{s}...M^{\sigma_L}. \nonumber
\end{align}
Here $S_{s,\sigma_i}$ is the swap operator and $M^{\sigma_1}...M^{s}M^{\sigma_{i}}...M^{\sigma_L}$ is the weight of the state $\ket{\sigma_1,...,s,\sigma_i,...,\sigma_L}$ in the MPS form with $s$ the index of the cavity mode site and $\sigma_i$ the index for the bosonic atoms. 
In Fig.~\ref{fig:mps3} we sketch how the swap gate acts on two MPS sites and changes their order. The swap gates are constructed from two Kronecker delta functions, each between indices of the same nature, but different sites, i.e. in Fig.~\ref{fig:mps3}(a) we have a Kronecker delta from the cavity index $s$ at the first site to the cavity index $s'$ at the second site (red curve) and a Kronecker delta from the atomic index $\sigma$ at the second site to the atomic index $\sigma'$ at the first site. The next step is the application of the swap gate onto the MPS wavefunction and obtaining a two-site tensor with swapped indices [Fig.~\ref{fig:mps3}(b)]. Finally an SVD decomposition is performed to restore the MPS structure. Thus, using the swap gates we can apply the operator $e^{-\frac{i\mathrm{d}t}{\hbar} \left(H_{\text{ac}}+H_{\text{c}}-\frac{i}{2}\hbar\Gamma a^\dagger a\right)}$ onto the wavefunction as a series of two-site gates, as depicted in Fig.~\ref{fig:mps4}. No additional error is introduced by the swap gates, except the SVD truncation error.

The implemented time-evolution method has an error of the order $\mathcal{O}(L\mathrm{d}t^2)$ at a certain final time $t$, stemming from the Trotter-Suzuki decomposition. However, the stochastic unravelling is only valid up to first order to $\mathrm{d}t$, such that we expect that this limits the choice of the time step. A detailed analysis of the contributing errors and improvements of the method using different time-evolution schemes will be performed in future. 

To improve the performance of our tMPS algorithm we take good quantum numbers into account, by noting that our system preserves the total number of bosonic atoms.

\subsection{\label{sec:conv}Numerical convergence}

\begin{figure}[hbtp]
\centering
\includegraphics[width=.5\textwidth]{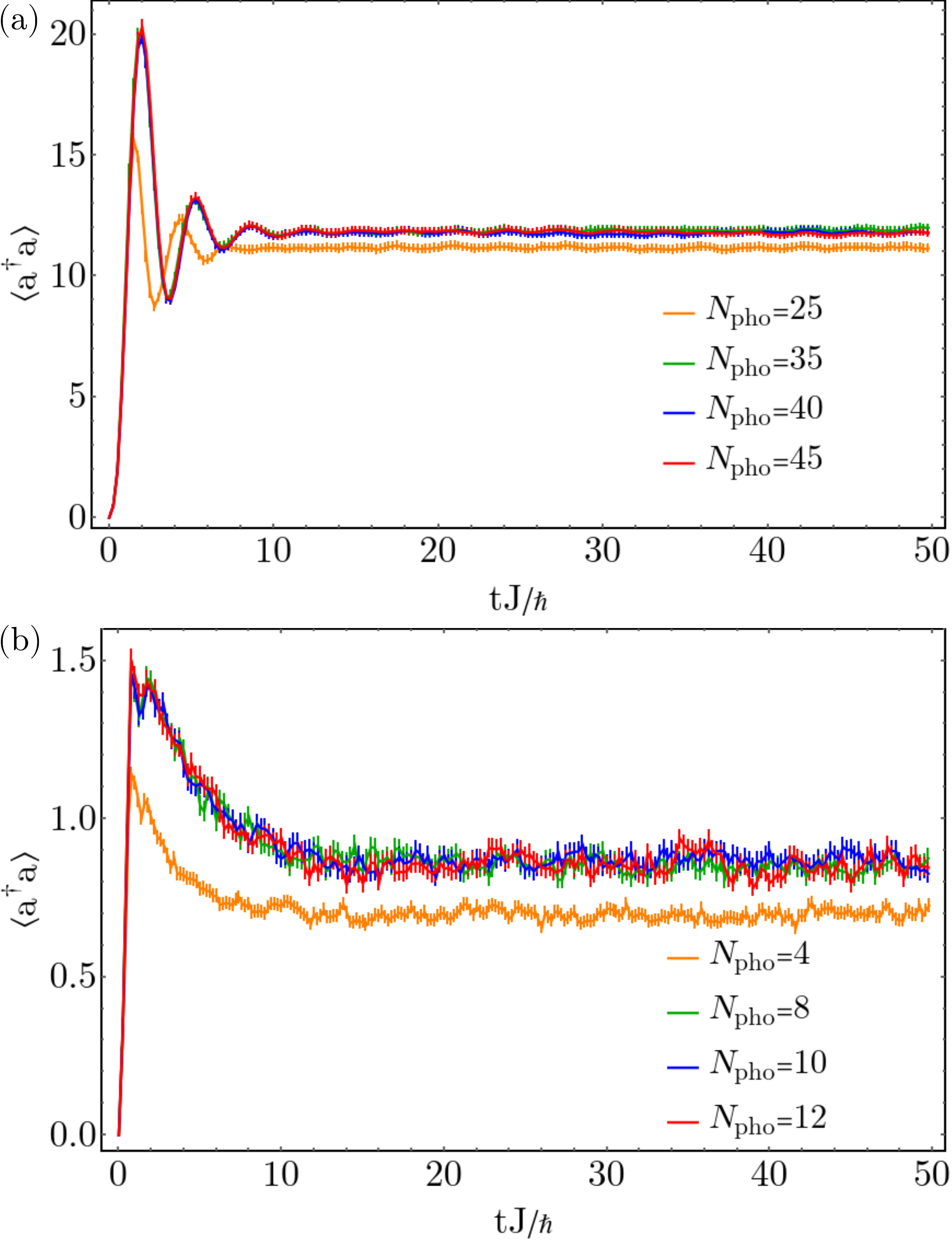}
\caption{
The time evolution of the photon number for different cut-offs of the photonic Hilbert space, $N_{\text{pho}}$. We present the behavior for two parameter sets, (a) $\hbar\Omega\sqrt{N}/J=3.35$, $\hbar\Gamma/J=1$,  $\mathrm{d}tJ/\hbar=0.0125$ and (b) $\hbar\Omega\sqrt{N}/J=4.47$, $\hbar\Gamma/J=10$ , $\mathrm{d}tJ/\hbar=0.01$. 
The error bars represent the standard deviation of the Monte Carlo average over 500 trajectories for (a), and 750 trajectories for (b). We use $L=10$, $N=5$, $\hbar\delta/J=2$, $U/J=2$, and $\epsilon=10^{-12}$. }
\label{fig:convergence1}
\end{figure}

\begin{figure}[hbtp]
\centering
\includegraphics[width=.5\textwidth]{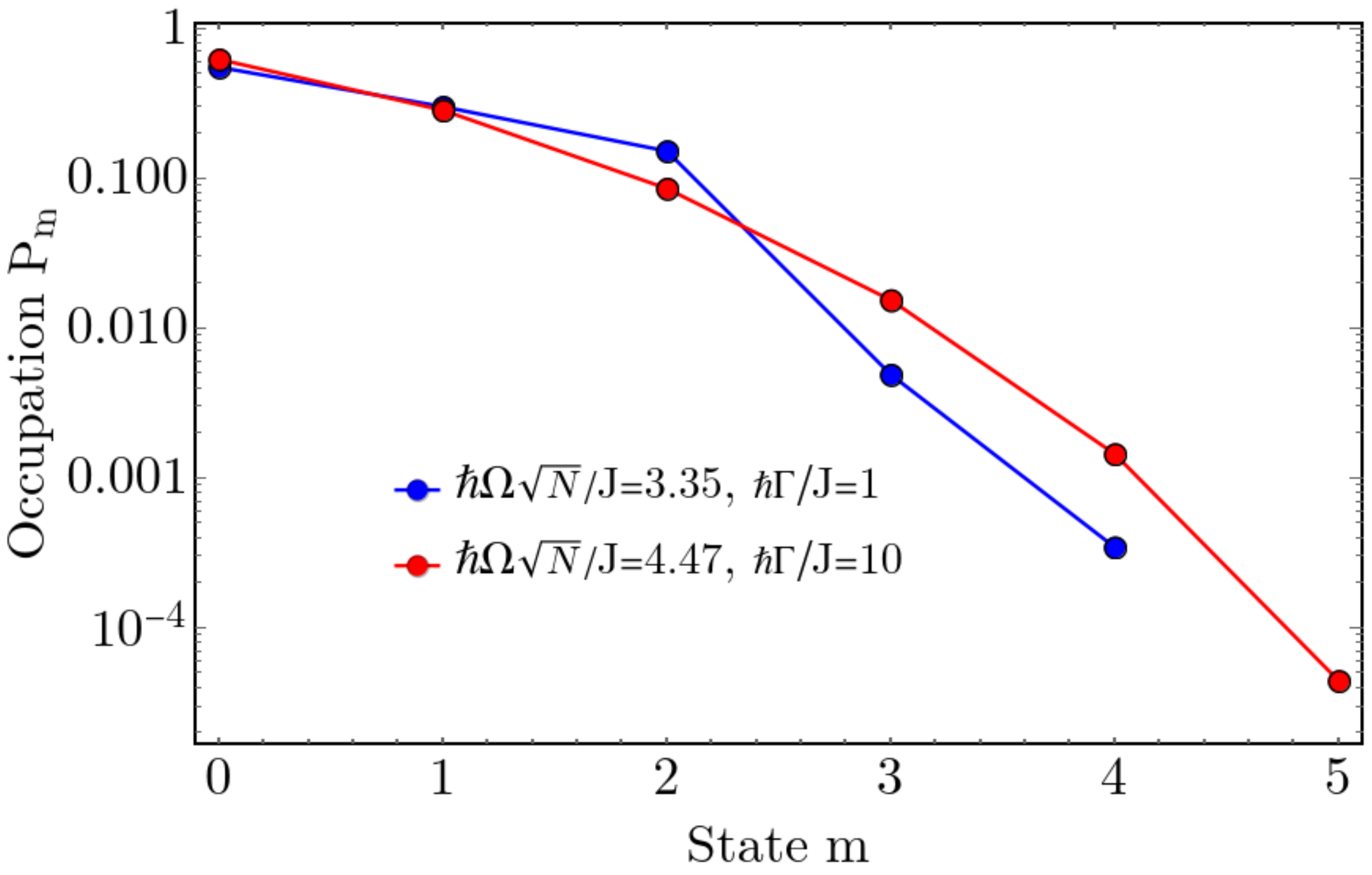}
\caption{
The boson number distribution, $P_m=\text{tr}\left(\bra{m}\rho\ket{m}\right)$, in the middle of the chain, for site $5$, at $tJ/\hbar=49.75$. We present the behavior for two parameter sets, (red dots) $\hbar\Omega\sqrt{N}/J=3.35$, $\hbar\Gamma/J=1$,  $\mathrm{d}tJ/\hbar=0.0125$, $N_\text{pho}=40$ and (blue dots) $\hbar\Omega\sqrt{N}/J=4.47$, $\hbar\Gamma/J=10$ , $\mathrm{d}tJ/\hbar=0.01$, $N_\text{pho}=10$. We use $L=10$, $N=5$, $\hbar\delta/J=2$, $U/J=2$, and $\epsilon=10^{-12}$,}
\label{fig:boscut}
\end{figure}

The convergence of the numerical method is controlled by several parameters. The stochastic unravelling of the master equation with quantum trajectories requires an averaging of a sufficiently large number of trajectories. Additionally, the time step $\mathrm{d}t$ must be chosen small enough in order to avoid the occurence of multiple jumps in one time step. The next source of error comes from the Trotter-Suzuki decomposition of the time evolution operator. Again this require that the time-step $\mathrm{d}t$ is small enough. Finally, we introduce an additional error by representing our wavefunctions as a matrix product state with finite local and bond dimensions. This implies a cut-off, $N_\text{pho}$, of the local Hilbert space of the photons and in some situations also for the bosonic atoms. The procedure to dynamically adjust the cut-off, $N_\text{pho}$, will be presented in Sec.~\ref{sec:photon}. Additionally, the introduction of the finite bond dimension using SVD leads to a so-called truncation error. To control the bond dimension of the MPS we impose a truncation error goal $\epsilon$, thus, in each compression step, after the application of a time evolution or swap gate onto the MPS, the number of states kept is such that the truncation error is smaller than $\epsilon$. Let us note that as in the case of the time-dependent MPS \cite{DaleyVidal2004, WhiteFeiguin2004}, the arising errors are not independent and therefore, a careful analysis needs to be performed.

In the following we discuss typical values of the convergence parameters and their influence on the results. The discussion of the truncation error and the von Neumann entropy of the trajectories is presented in Sec. \ref{sec:ent}.

\begin{figure*}[hbtp]
\centering
\includegraphics[width=.95\textwidth]{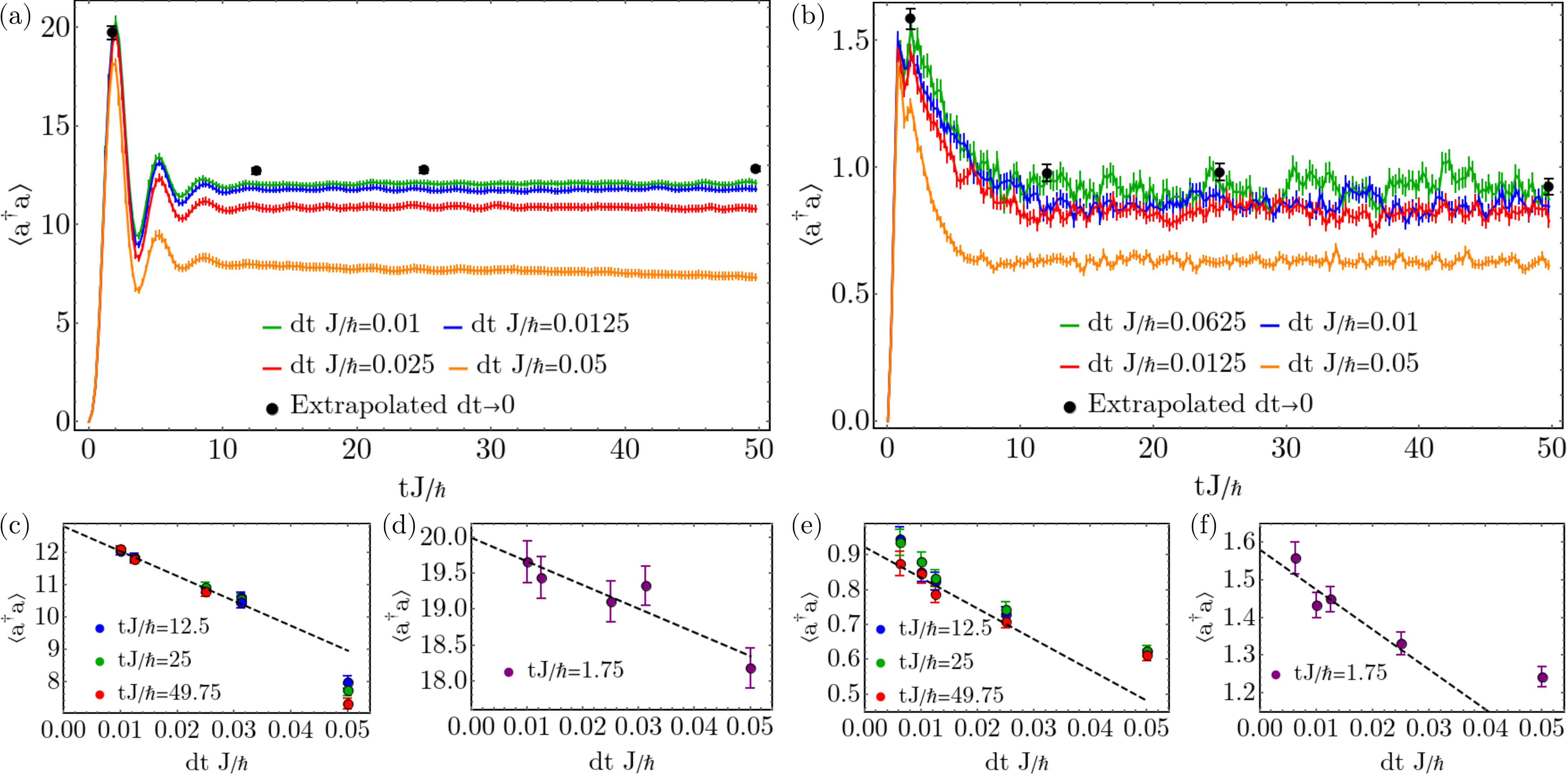}
\caption{
(a)-(b) The time evolution of the photon number for different time steps $\mathrm{d}t$. The black dots represents the extrapolated value in the limit $\mathrm{d}tJ\to 0$ at $tJ/\hbar\in\{1.75, 12, 25, 49.75\}$. 
(c)-(f) Convergence of the photon number with the time step at several chosen times. The dashed line represents a linear fit of the dependence on $\mathrm{d}tJ$, for (c) and (e) the fit is done for the data taken at $tJ/\hbar=49.75$.
We present the behavior for two parameter sets, (a), (c), (d) $\hbar\Omega\sqrt{N}/J=3.35$, $\hbar\Gamma/J=1$, $N_\text{pho}=40$ and (b), (e), (f) $\hbar\Omega\sqrt{N}/J=4.47$, $\hbar\Gamma/J=10$ and $N_\text{pho}=10$. The error bars represent the standard deviation of the Monte Carlo average over  500 trajectories for (a), (c), (d) and 750 trajectories for (b), (e), (f). We use $L=10$, $N=5$, $\hbar\delta/J=2$, $U/J=2$, and $\epsilon=10^{-12}$.}
\label{fig:convergence2}
\end{figure*}

{\it Stochastic error.} 
We estimate the error of having a finite number of quantum trajectories included in the Monte Carlo average by computing the standard deviation of the mean for the measured expectation value of an operator $\mathcal{E}$ 
\begin{align}
\label{eq:std}
& \sigma\left(\mathcal{E}(t)\right)=\sqrt{\frac{1}{R(R-1)}\sum_{r=1}^R \left(\bra{\psi_r (t)}\mathcal{E}\ket{\psi_r (t)}-\langle\langle\mathcal{E}\rangle\rangle\right)^2},
\end{align}
where $R$ is the total number of samples, $\ket{\psi_r (t)}$ the time evolved wavefunction of the trajectory labelled by $r$, and $\langle\langle\mathcal{E}\rangle\rangle$ the statistical average over all quantum trajectories.
For the numerical data presented, we show this error with the curves when averages are presented.  
Typically, we average over at least $500$ trajectories, which ensures that for the physical parameters considered the relative error in the expectation value of the photon number is smaller than $3\%$  (see for example Fig.~\ref{fig:convergence1} upper panel). For the cases when the photon number is small, $\langle a^\dagger a \rangle\lesssim1$, either at small coupling $\Omega$ or large dissipation strengths $\Gamma$, we average over $750$ trajectories to obtain the same relative error, as the fluctuations have a greater influence (Fig.~\ref{fig:convergence1} lower panel).

{\it Cut-off of the dimension of the local Hilbert spaces.} 
The dimension of the local Hilbert space for the photon can be infinite and thus, a cut-off for its dimension is needed in the numerical implementation.
In the following we will denote the cut-off $N_{\text{pho}}$, referring to the maximal number of photons that we can capture and noting that we use  $N_{\text{pho}}+1$ Fock states as we also have the vacuum state.
In order to identify more clearly the influence of the cut-off on the results we used a fixed cut-off for the photonic site in this section. However, in Sec.~\ref{sec:photon} we present a more efficient approach by implementing an adaptive photonic local dimension, since the required cut-off varies considerably in time and with the trajectories.
Examples with a fixed cut-off with all other parameters fixed are shown in Fig.~\ref{fig:convergence1}. For a given set of parameters we observe that above a certain value of the cut-off $N_{\text{pho}}$ the average value of the photon number is only slightly varying with increasing the cut-off. In particular for the presented situation its variation for $N_{\text{pho}}\ge 35$ becomes lower that the error bars of the Monte Carlo averaging. However, choosing a too low cutoff e.g.~$N_{\text{pho}}\le 25$ ($N_{\text{pho}}\le 4$) in Fig.~\ref{fig:convergence1}, leads to misleading results for both the time-evolution and the reached long time values. Note, that even though the long time value lies around 12 (below 1)  photons, taking double this value for the cutoff i.e. 25 (4) is not sufficient. The required cut-off depends very much on the physical parameters. Therefore, one needs to consider each case separately, which can result in very different values for the cut-off.

Since we consider bosonic atoms, the maximal possible local dimension for the atomic sites equals the total atom number plus one for the possibility to have an empty site. We found that often this very large local dimension is needed, which also strongly restricts the total number of atoms which can be efficiently considered. However, in some situations a reduced dimension of the local bosonic site can be taken. In Fig.~\ref{fig:boscut} we compare the occupations of each bosonic number state for a site in the middle of the chain. We can observe that for the parameter set with $\hbar\Gamma/J=1$ the occupations of the states with a large boson number are a few times smaller than for the parameter set with $\hbar\Gamma/J=10$. We note that the other sites in the chain have even smaller occupations of the states with three or four bosons, for $\hbar\Gamma/J=1$. Therefore,  a maximal local dimension of five instead of six is sufficient in the considered case. 

{\it Influence of the time step.}
The dependence of the results on the value of the time step is more involved. This is due to the fact that the time step controls both the convergence of the stochastic sampling process and the Trotter-Suzuki decomposition. Further, as in the normal time-dependent MPS, the time step interplays with the truncation error in a non-trivial fashion, since a smaller time-step requires more truncations and therefore results in an increased truncation error \cite{DaleyVidal2004}. Therefore, the values used needs to be adjusted very carefully depending not only on the physical parameters but also on the convergence parameters of the model.

\begin{figure}[hbtp]
\centering
\includegraphics[width=.5\textwidth]{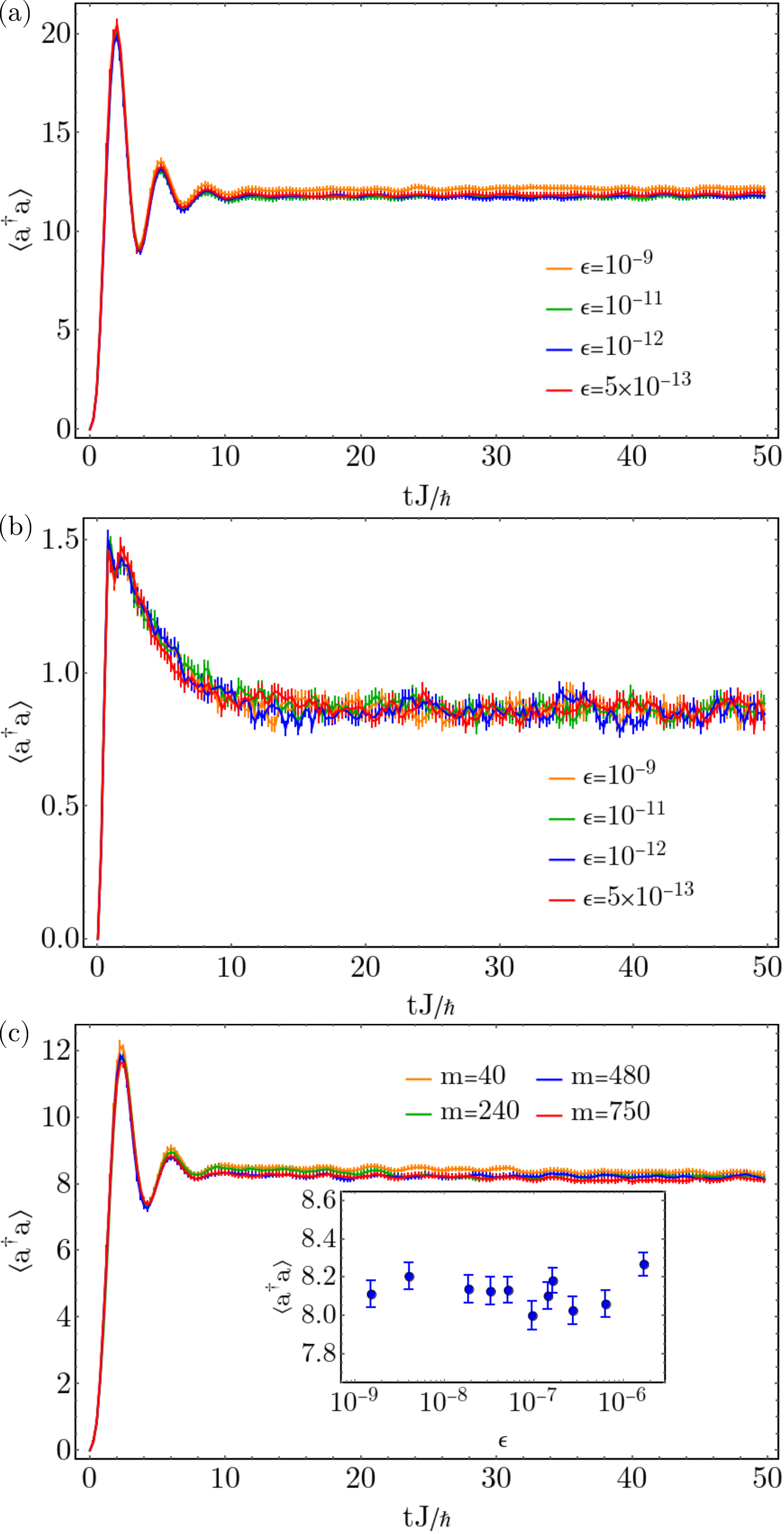}
\caption{
The time evolution of the photon number for different truncation errors, $\epsilon$ and bond dimensions, $m$. We present the behavior for three parameter sets, (a) $\hbar\Omega\sqrt{N}/J=3.35$, $\hbar\Gamma/J=1$, $L=10$ sites, $N=5$ particles, (b) $\hbar\Omega\sqrt{N}/J=4.47$, $\hbar\Gamma/J=10$, $L=10$ sites, $N=5$ particles and (c) $\hbar\Omega\sqrt{N}/J=2.46$, $\hbar\Gamma/J=1$, $L=14$ sites, $N=7$ particles. We use $\hbar\delta/J=2$ and $U/J=2$.
In the inset of (c) the photon number is taken at $tJ/\hbar=49.75$.
The error bars represent the standard deviation of the Monte Carlo average. The numerical parameters used in the tMPS method are the following: the time step is $\mathrm{d}tJ/\hbar=0.0125$ in (a) and (c), $\mathrm{d}tJ/\hbar=0.01$ in (b), and the cut-off of the local dimension for the photon mode is $N_{\text{pho}} = 40$ in (a), $N_{\text{pho}} = 10$ in (b) and dynamically adapted in (c) (see Sec.~\ref{sec:photon}). The Monte-Carlo average contains 500 trajectories for (a) and (c), and 750 trajectories for (b).
 }
\label{fig:convergence3}
\end{figure}

In Fig.~\ref{fig:convergence2} we show an example of the variation obtained fixing all parameters beside the time step $\mathrm{d}t$. A relatively rapid convergence is seen using time-steps between $\mathrm{d}tJ/\hbar=0.01-0.05$. In particular, the convergence is in agreement with the expected linear behaviour in the time-step $\textrm{d}t$ which suggests a well justified extrapolation method. 
For small values of $\Gamma$, here $\hbar \Gamma/J=1$, the error induced by the time-step remains larger than the error of the statistical error. The extrapolated value lies a bit above the shown results at a finite time step. In contrast for the case of large $\Gamma$, the statistical error is dominating the results and the extrapolated result lies within the statistical error bars of the smallest time steps. 

\begin{figure}[hbtp]
\centering
\includegraphics[width=.5\textwidth]{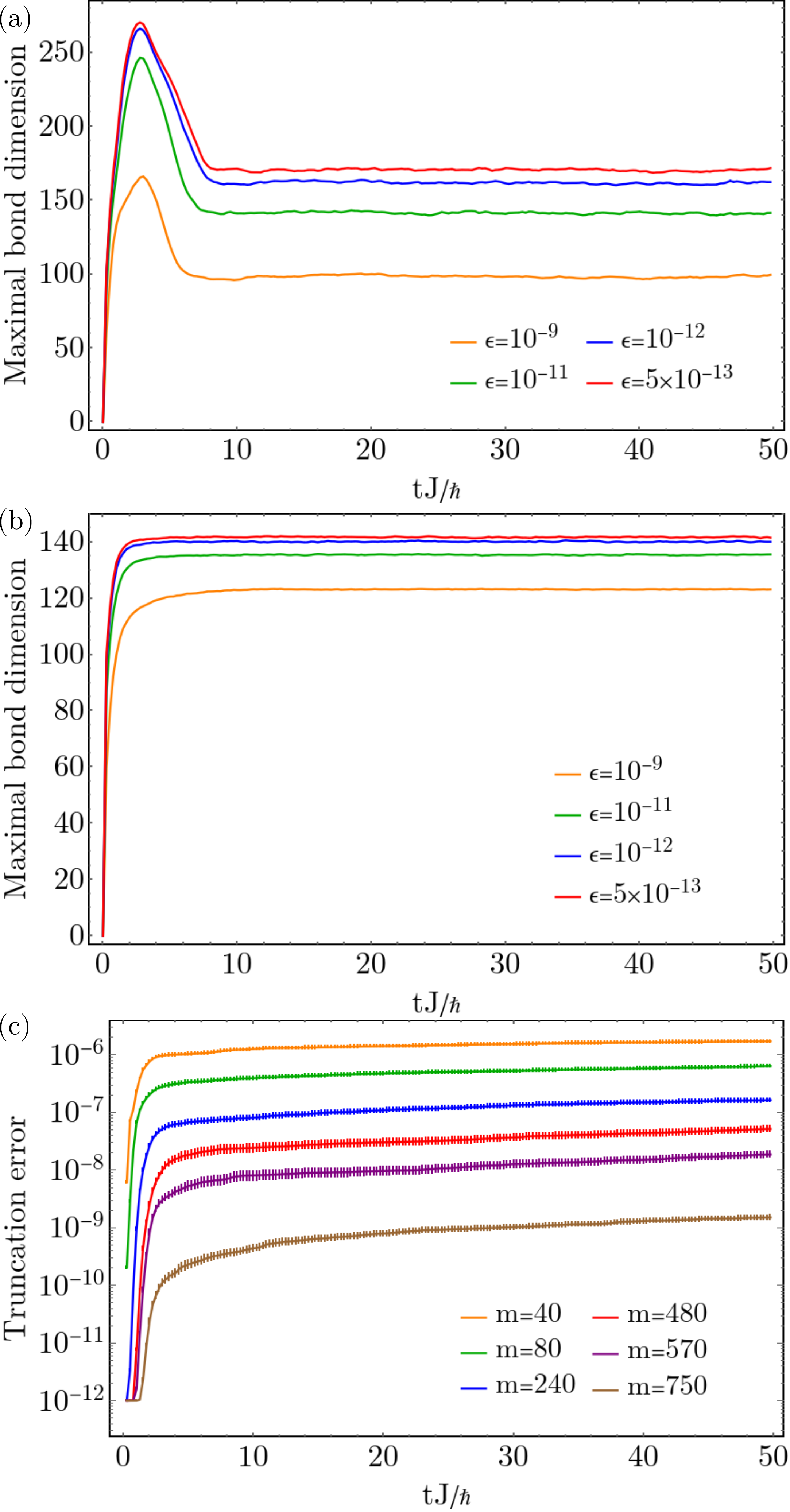}
\caption{
The time evolution of the maximal bond dimension for different truncation errors, $\epsilon$ (a), (b), and the time evolution of the truncation error  $\epsilon$ for different bond dimensions, $m$. We present the behavior for three parameter sets, (a) $\hbar\Omega\sqrt{N}/J=3.35$, $\hbar\Gamma/J=1$, $L=10$ sites, $N=5$ particles, (b) $\hbar\Omega\sqrt{N}/J=4.47$, $\hbar\Gamma/J=10$, $L=10$ sites, $N=5$ particles, and (c) $\hbar\Omega\sqrt{N}/J=2.46$, $\hbar\Gamma/J=1$, $L=14$ sites, $N=7$ particles. We use $\hbar\delta/J=2$ and $U/J=2$.
The numerical parameters used in the tMPS method are the following: the time step is $\mathrm{d}tJ/\hbar=0.0125$ in (a) and (c), and $\mathrm{d}tJ/\hbar=0.01$ in (b), and the cut-off of the local dimension for the photon mode is $N_{\text{pho}}=40$ in (a), $N_{\text{pho}}=10$ in (b) and dynamically adapted in (c) (see Sec.~\ref{sec:photon}). The Monte-Carlo average contains 500 trajectories for (a) and (c), and 750 trajectories for (b).
 }
\label{fig:convergence4}
\end{figure}

\begin{figure}[!hbtp]
\centering
\includegraphics[width=.5\textwidth]{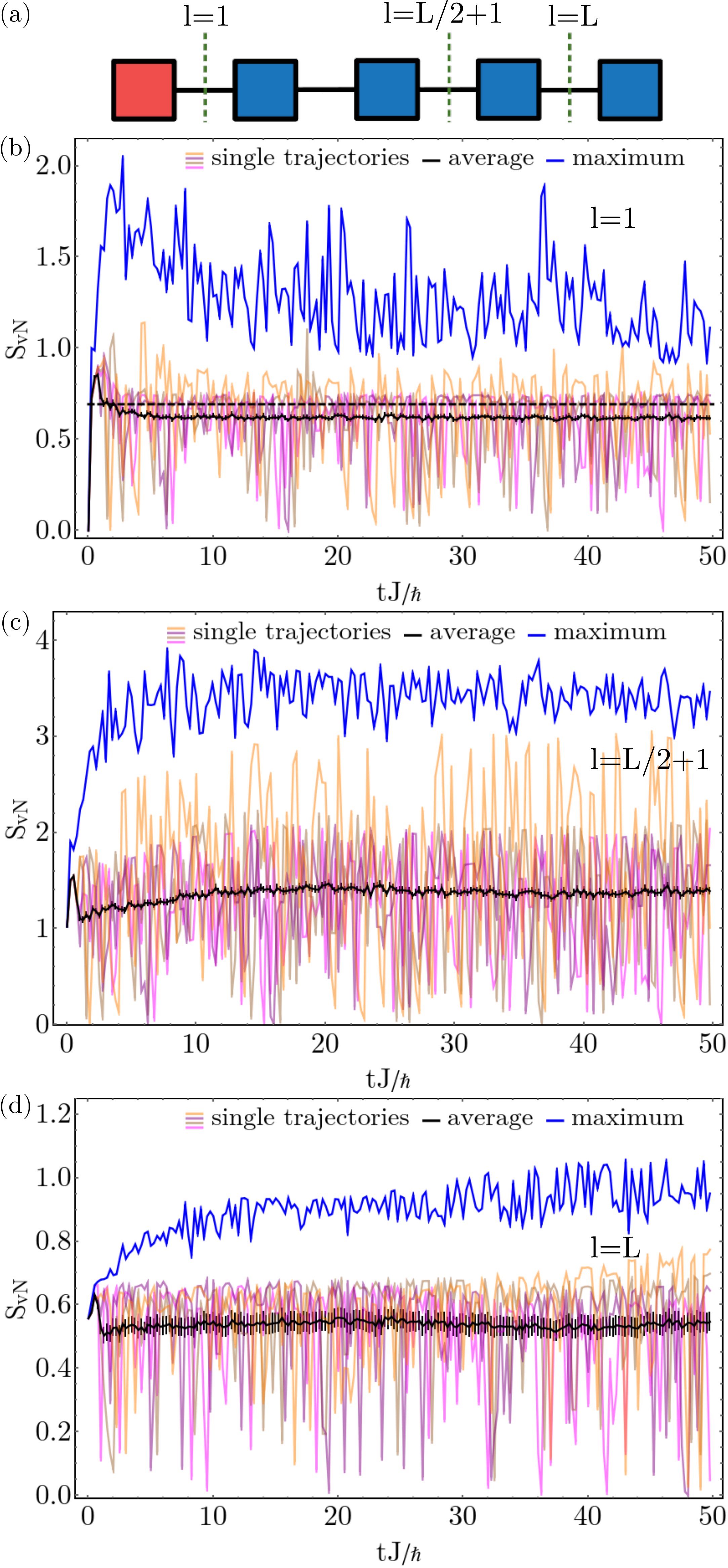}
\caption{
(a) The graphical representation of the MPS structure denoting the bonds for which the von Neumann entropy was computed, for a bipartition between the cavity site and the atomic sites with $l=1$, a bipartition in the middle of the atomic chain, with one half also containing the cavity site $l=L/2+1$, and a bipartition between the last atomic site and the rest of the chain, $l=L$.
(b)-(d) The time evolution of the von Neumann entropy, $S_\text{vN}$, of a few single trajectories, the Monte Carlo average and the maximum value over different trajectories for (b) $l=1$, (c) $l=L/2+1$ and (d) $l=L$.
The parameters used are $L=10$, $N=5$, $\hbar\delta/J=2$, $U/J=2$, $\hbar\Omega\sqrt{N}/J=3.35$ and $\hbar\Gamma/J=1$. The Monte-Carlo averages contain at least 500 trajectories.
 }
\label{fig:ent1}
\end{figure}

\begin{figure}[!hbtp]
\centering
\includegraphics[width=.5\textwidth]{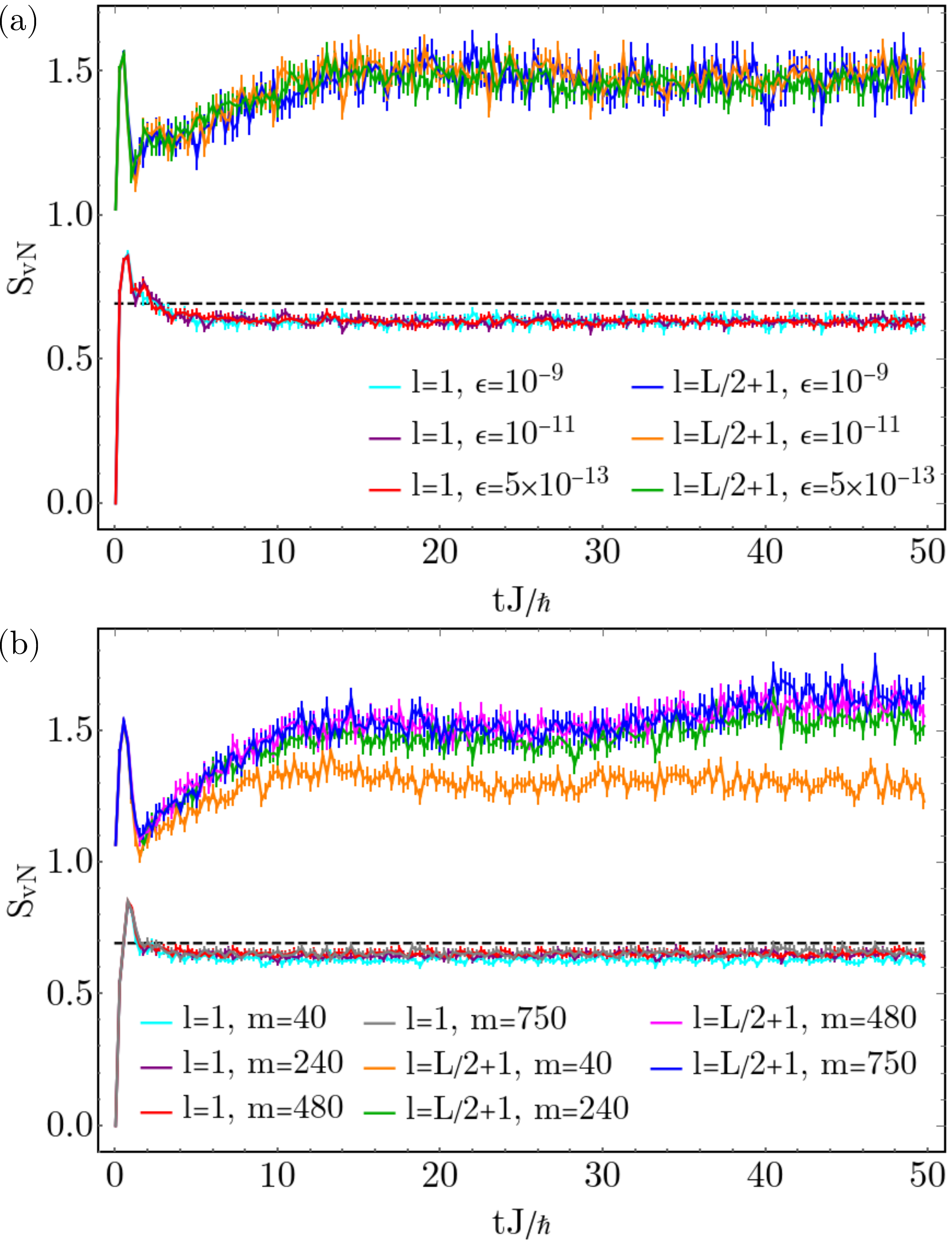}
\caption{
The time evolution of the von Neumman entropy $S_\text{vN}$ for $l=1$, $l=L/2+1$ and (a) different truncation errors  and (b) bond dimensions 
The parameters used are $\hbar\delta/J=2$, $U/J=2$, (a) $\hbar\Omega\sqrt{N}/J=3.35$, $\hbar\Gamma/J=1$, $L=10$ sites, $N=5$ particles and (b) $\hbar\Omega\sqrt{N}/J=2.46$, $\hbar\Gamma/J=1$, $L=14$ sites, $N=7$ particles. The Monte-Carlo averages contain at least 500 trajectories.
 }
\label{fig:ent11}
\end{figure}

\begin{figure}[hbtp]
\centering
\includegraphics[width=.5\textwidth]{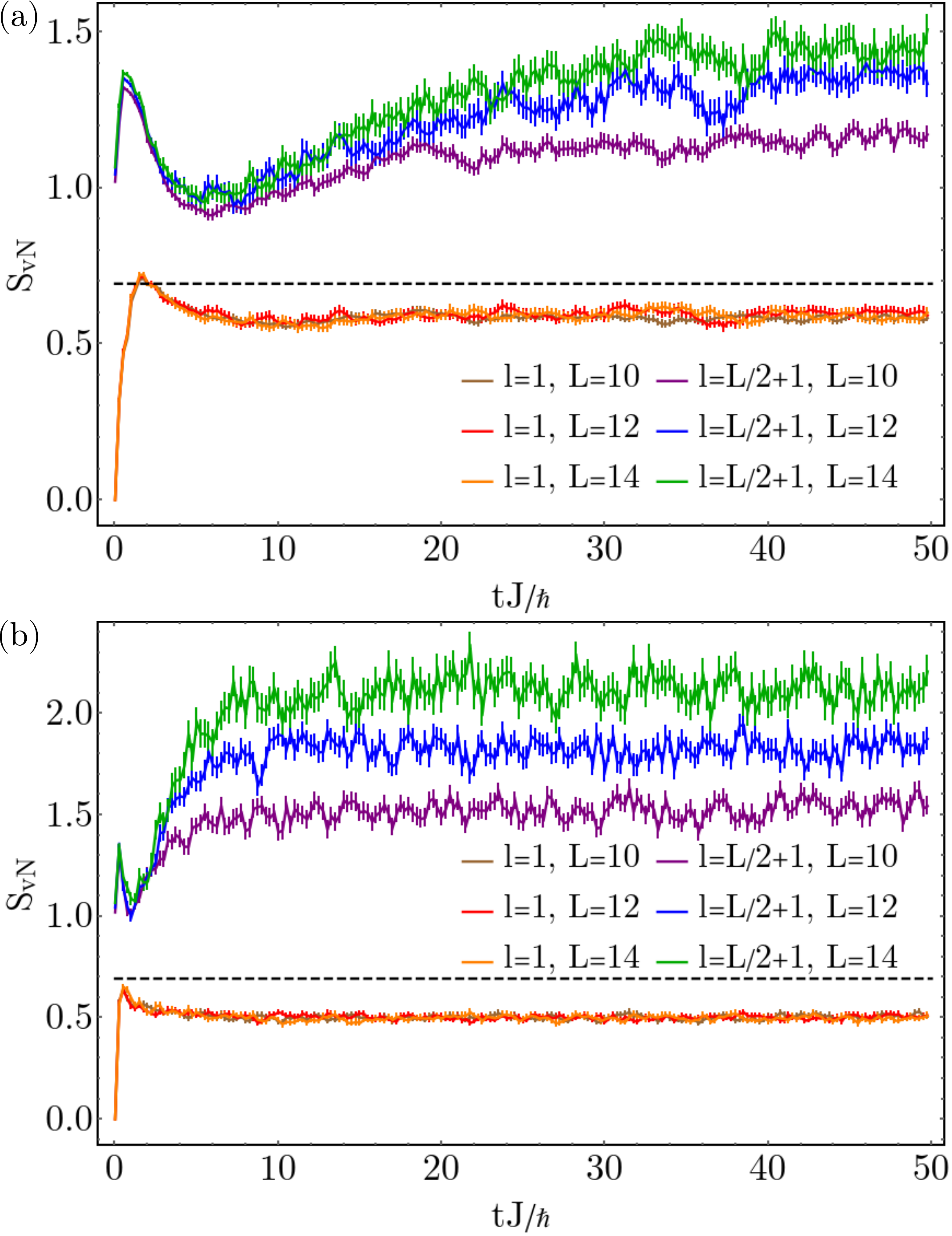}
\caption{
The time evolution of the von Neumman entropy, $S_\text{vN}$, of the Monte Carlo average for different system size, $L\in\{10,12,14\}$, for two bipartitions $l=1$ and $l=L/2+1$.
The parameters used are $N=L/2$, $\hbar\delta/J=2$, $U/J=2$, (a) $\hbar\Omega\sqrt{N}/J=1.6$ and $\hbar\Gamma/J=1$, (b) $\hbar\Omega\sqrt{N}/J=4.47$ and $\hbar\Gamma/J=13$.
The Monte-Carlo averages contain at least 500 trajectories.
 }
\label{fig:ent2}
\end{figure}

\subsection{\label{sec:ent}Entanglement of quantum trajectories}
One of the most important convergence parameters is the bond dimension, $m$, used within the SVD compressions. One measure of this is the truncation error $\epsilon$, the sum of the neglected eigenvalues of the reduced density matrix in the SVD compression.  Another measure of the decay of the eigenvalues is the von Neumann entropy, $S_\text{vN}$. We analyze in the following the behavior of these two quantities for different parameters.

We first look at the dependence on the truncation error $\epsilon$ of the singular value decomposition performed in the time evolution gates and swap gates shown in Fig.~\ref{fig:convergence3}. As the truncation error is chosen relatively small for $L=10$ [Fig.~\ref{fig:convergence3}(a)-(b)], the results only weakly depend on the value of the maximal truncation error $\epsilon$. In the case of $L=14$ [Fig.~\ref{fig:convergence3}(c)], where we control the truncation error by fixing the used bond dimension, we can observe that the obtained photon numbers are consistent with each other, except for the case with $m=40$, which corresponds to $\epsilon\approx 10^{-6}$ at $tJ/\hbar=49.75$.
In particular, the deviations are of the order to the statistical error. Thus, we are confident that a truncation error of $\lesssim 10^{-7}$ provides an accurate description of the considered states in the matrix product form. 

One can also monitor the maximal bond dimension needed in the MPS representation in order to achieve the set truncation error goal, as depicted in Fig.~\ref{fig:convergence4} (a)-(b), or the largest truncation error obtained for a fixed bond dimension in Fig.~\ref{fig:convergence4}(c). The maximal bond dimension increases considerably with lowering the truncation error. However, for the smallest chosen truncation errors the maximal bond dimension saturates. We can observe that the bond dimension needed to describe the system is between 100 and 300 even for a system of size $L=10$. For a system of size $L=14$ [Fig.~\ref{fig:convergence4}(c)] one needs to increase the bond dimension with more than an order of magnitude, from $m=40$ to $m=750$, in order to decrease the truncation error from $\epsilon \sim 10^{-6}$ to $\epsilon \sim 10^{-9}$ at long times.

In the following we turn to the von Neumann entropy of the quantum trajectories to monitor the coupling of the photonic and atomic sectors and the correlations within the atomic chain. We note that the entanglement entropy of the quantum trajectories is not a direct measure of the entanglement present in the density matrix resulting from the Monte Carlo averaging process. However, $S_\text{vN}$ provides valuable information about how well our MPS method captures the entanglement present in the trajectories. 

In the following, we consider three different bipartitions of the MPS, as depicted in Fig.~\ref{fig:ent1}(a), between the cavity site and the rest of the atomic chain, bond $l=1$, in the middle of the atomic chain, where one half also contains the cavity site, bond $l=L/2+1$, and the last bond $l=L$. This is motivated by our finding that the maximum of $S_\text{vN}$ throughout the atomic chain occurs at the bond $l=L/2+1$, and the final atomic side is the furthest apart from the cavity site.
In Figs.~\ref{fig:ent1}(b)-(d) we present the time evolution of the entropy for the Monte Carlo average, the maximum entropy of the sampled quantum trajectories and for a few single trajectories, for the three considered bipartitions. We observe that for all bipartitions $S_\text{vN}$ saturates to a finite value in time, both for the average and maximum values. 

In the long time limit, the von Neumann entropy takes finite values for all bipartitions and parameter sets considered. In Fig.~\ref{fig:ent1}(b) we see that at low dissipation strength the average entropy computed between the photon mode and the atoms ($l=1$) becomes close to $\log(2)$ at long times. This signals a coherent superposition of two states which we attribute to a superposition of states with a different sign of the photon field \cite{HalatiKollath2020}. The value of the entanglement within the chain is larger which points to the contribution of several states in the superposition. 

In Fig.~\ref{fig:ent11}(a) it can be seen that the values of $S_\text{vN}$ only change within the Monte Carlo averaging uncertainty for all considered truncation errors for $L=10$. Thus, we can be confident that our method captures the dynamics of our system correctly up to long times. In the case of $L=14$, Fig.~\ref{fig:ent11}(b), we can observe that the von Neumann entropy, $S_\text{vN}$, computed in the middle of the chain is accurately described for a bond dimension larger than $m\geq 240$. This bond dimension corresponds to $\epsilon\approx 10^{-7}$ at $tJ/\hbar=49.75$ [Fig.~\ref{fig:convergence4}(c)].
As most results in this work were computed with a truncation error goal of $10^{-12}$, the perspective of pushing the numerical simulations towards larger systems and longer times by considering a larger truncation error can be envisioned. 

In Fig.~\ref{fig:ent2} we computed the entropy for different system sizes, for two different parameter sets. We observe that in both cases the entanglement present in the quantum trajectories between the photon mode and the atoms seems stable with the respect to the system size. This further supports the claim that a coherent superposition of two system size independent states contribute, as it is the case for the states with the different sign of the photon field \cite{HalatiKollath2020}. For the bond $l=L/2+1$ we see that the value at which the entropy saturates increases with the system size, indicating that the system might be in a gapless phase or that the system size is not yet long enough to cover the correlation length of the gapped state.

\begin{figure}[hbtp]
\centering
\includegraphics[width=.48\textwidth]{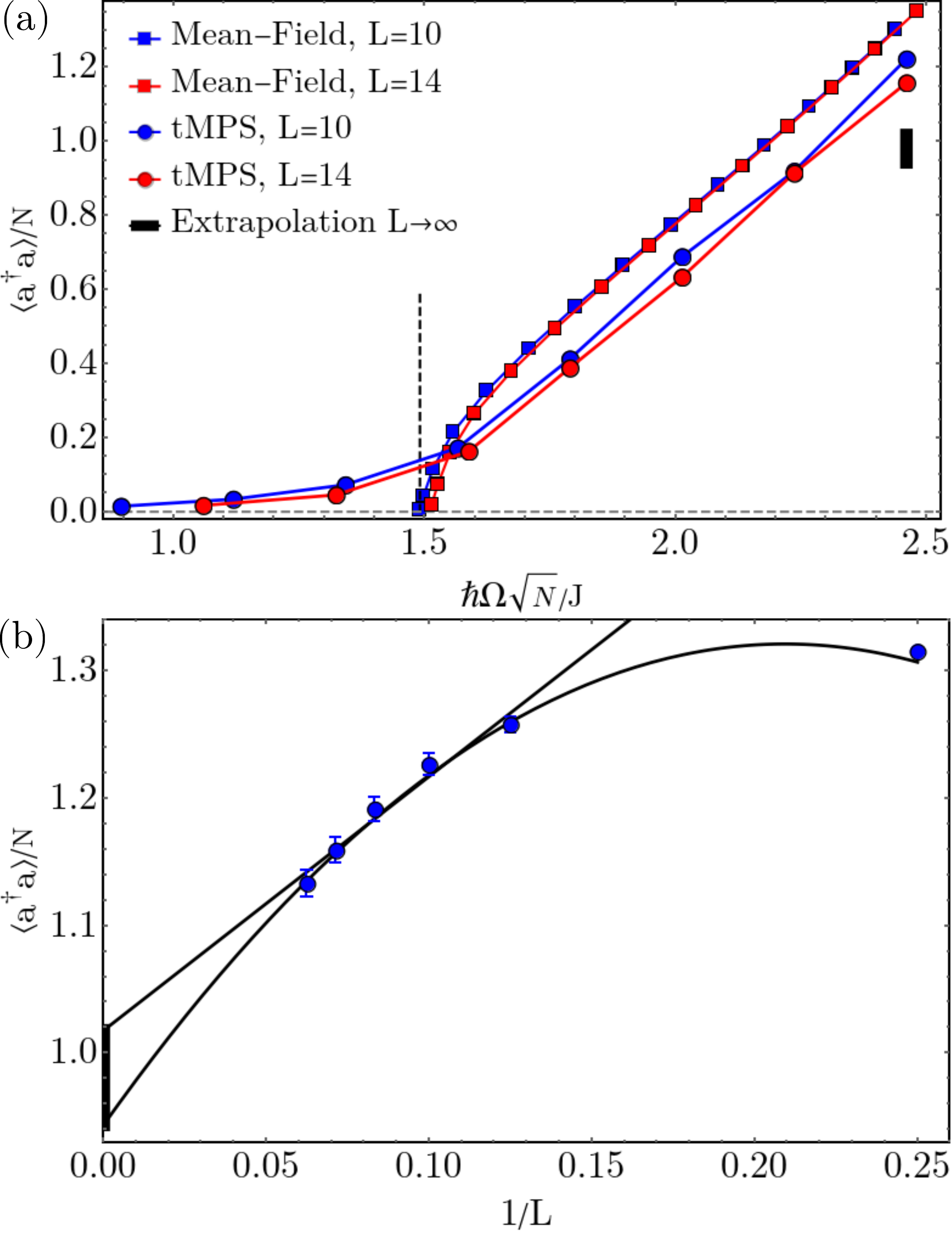}
\caption{
(a) The scaled photon number, $\langle a^\dagger a\rangle/N$, as a function of the scaled atoms-cavity coupling $\hbar\Omega\sqrt{N}/J$ for $L\in \{10,14\}$.  We compare our numerical results with the mean field approach. The dashed vertical line marks the self-organization threshold as obtained from the mean-field approach for $L=10$ sites. The black vertical line represents the interval between the two possible extrapolations depicted in (b).
(b) The scaled photon number, $\langle a^\dagger a\rangle/N$, as a function of the inverse system size, $1/L$, for $\hbar\Omega\sqrt{N}/J=2.46$. The black curves represent a linear and a quadratic fit.
The parameters are $n=N/L=1/2$, $\hbar\delta/J=2$, $U/J=2$, $\hbar\Gamma/J=1$, and $tJ/\hbar=49.75$. The point for $L=4$ is obtained by the exact diagonalization of Eqs.~(\ref{eq:Lindblad})-(\ref{eq:Hamiltonian}).
The numerical parameters used in the tMPS method are the truncation error $\epsilon = 10^{-12}$ for $L=10$ and $\epsilon = 10^{-9}$ for $L>10$, the cut-off of the local dimension for the photon mode between $10$ and $25$, adapted to the average photon number, and $\mathrm{d}tJ/\hbar=0.0125$. The Monte-Carlo averages contain at least 500 trajectories.
 }
\label{fig:conv_n}
\end{figure}

\subsection{\label{sec:finite}Finite size effects}

As we have seen in the previous subsections, at long times the considered quantities have become almost constant in time. Typically, such a regime is reached long before $tJ/\hbar\approx 50$, as shown in the time evolution plots, for example Fig.~\ref{fig:convergence1} for the photon number, or Fig.~\ref{fig:ent1} for the von Neumann entropy. Therefore, in this subsection we compare the values at late times for different system sizes, as we interpret these as very good approximations of the steady state values. 

In order to evaluate the finite size effects we analyze how the transition from the normal state to the self-organized state takes place for different system sizes. In Fig.~\ref{fig:conv_n}(a) we scale the photon number and the atoms-cavity coupling with the number of particles. For a comparison we show both the mean field and the numerically exact tMPS method results. Both show only small deviations with increasing the system size. 
In particular, in the mean field results the transition to the self-organizes phase starts later and becomes steeper with increasing system size. In the tMPS results, the rise of the photon number also seems to occur for a bit larger scaled pump strength and the scaled photon number is slightly lower for $L=14$. 

In Fig.~\ref{fig:conv_n}(b) we computed the scaled photon number for multiple system size in order to perform an extrapolation in the thermodynamic limit, $L\to\infty$. We note that the point at $L=4$ is obtained by the exact diagonalization of Eqs.~(\ref{eq:Lindblad})-(\ref{eq:Hamiltonian}) and taking the expectation value of the photon number in the steady state, the points for $L\geq 8$ were obtained using the tMPS method.
We observe that the photon number is monotonically decreasing for larger $L$, but the functional form of the system size dependence is not unambiguous, as both a linear fit of the points with $L\geq 8$ and a quadratic fit of all points can describe the behavior.  
However, as the extrapolated values are not far from the finite size results and they seems to move away from the mean field results (see black vertical line in Fig.~\ref{fig:conv_n}), it leads us to the expectation that our main findings will remain valid for large systems. 

To further support this, in Fig.~\ref{fig:ae}(b) the scaled photon number is plotted as a function of the dissipation strength for large dissipation strengths. The numerical results are compared with the many body adiabatic elimination results, as the state $\rho_\text{mix}$ can be evaluated for any system size. We observe that the agreement is very good at large dissipation strengths for all values of $L$ and $\Gamma$ considered. The scaled photon number is slightly decreasing for larger systems sizes in both approaches. This is consistent with the expected vanishing of the scaled photon number in the thermodynamic limit behavior for $\rho_\text{mix}$, as shown in Sec.~\ref{sec:thermo}.

\subsection{\label{sec:photon}Dynamically adapted cut-off of the local dimension for the cavity site}

We have shown in Sec.~\ref{sec:conv} that the cut-off of the local Hilbert space for the photons is an important convergence parameter of the numerical simulations. In particular, a sufficiently large $N_\text{pho}$ --typically about triple the average value-- is required to capture the dynamics correctly (Fig.~\ref{fig:convergence1}). Often during the time evolution the photon number varies considerably, as for example for the case in Fig.~\ref{fig:convergence1}(a). In this case at short times, $tJ/\hbar \lesssim  5$, there is a sudden increase of the number of photons in the cavity, larger than the value at late times. This suggests that we need a much larger $N_\text{pho}$ to accurately describe the photonic state at short times, than we would need at later times. Therefore, we decided to optimize our implementation by adapting the local dimension for the cavity site during the time evolution.

This improvement is similar in spirit with the recent developments regarding the time-dependent MPS methods with local basis optimization.
Refs. \cite{BrocktJeckelmann2015, StolppMeisner2020} apply the local basis optimization idea \cite{ZhangWhite1998} to the time evolution of the Holstein model of fermions locally coupled to phononic modes. In this approach one rotates the local Hilbert space adaptively into an optimized basis that can be truncated. In our case we find that even without changing the Fock basis for the photons we can dynamically adapt the number of states considered. The investigation whether an optimization of the photonic basis is further improving the algorithm is left for future implementations.

\begin{figure}[hbtp]
\centering
\includegraphics[width=.5\textwidth]{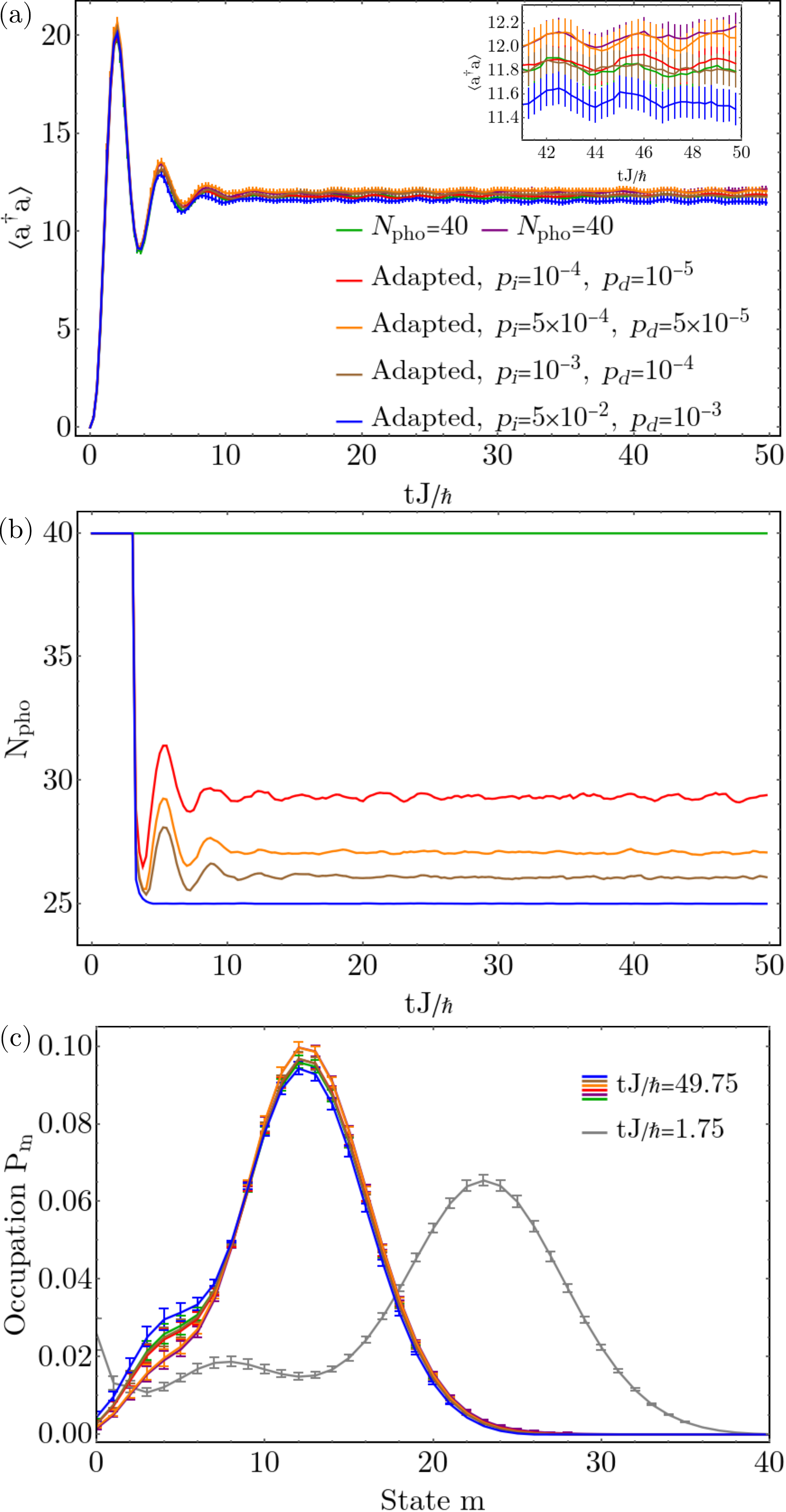}
\caption{
The time evolution of (a) the photon number, $\langle a^\dagger a\rangle$, and (b) $N_\text{pho}$. We compare the results corresponding to an adapted cut-off for different $p_i$ and $p_d$ with the Monte Carlo average of two  different sets of sampled trajectories with a fixed cut-off $N_{\text{pho}}=40$. 
(c) The photon number distribution, $P_m=\text{tr}\left(\bra{m}\rho\ket{m}\right)$, at $tJ/\hbar=49.75$ for the data presented in (a) and (b), and at $tJ/\hbar=1.75$ with a fixed cut-off. 
The parameters used are $L=10$, $N=5$, $\hbar\delta/J=2$, $U/J=2$, $\hbar\Omega\sqrt{N}/J=3.35$ and $\hbar\Gamma/J=1$, time step $\mathrm{d}tJ/\hbar=0.0125$ and the truncation error $\epsilon=10^{-12}$. The Monte-Carlo averages contain 500 trajectories. 
 }
\label{fig:adapt}
\end{figure}

\begin{figure}[!hbtp]
\centering
\includegraphics[width=.5\textwidth]{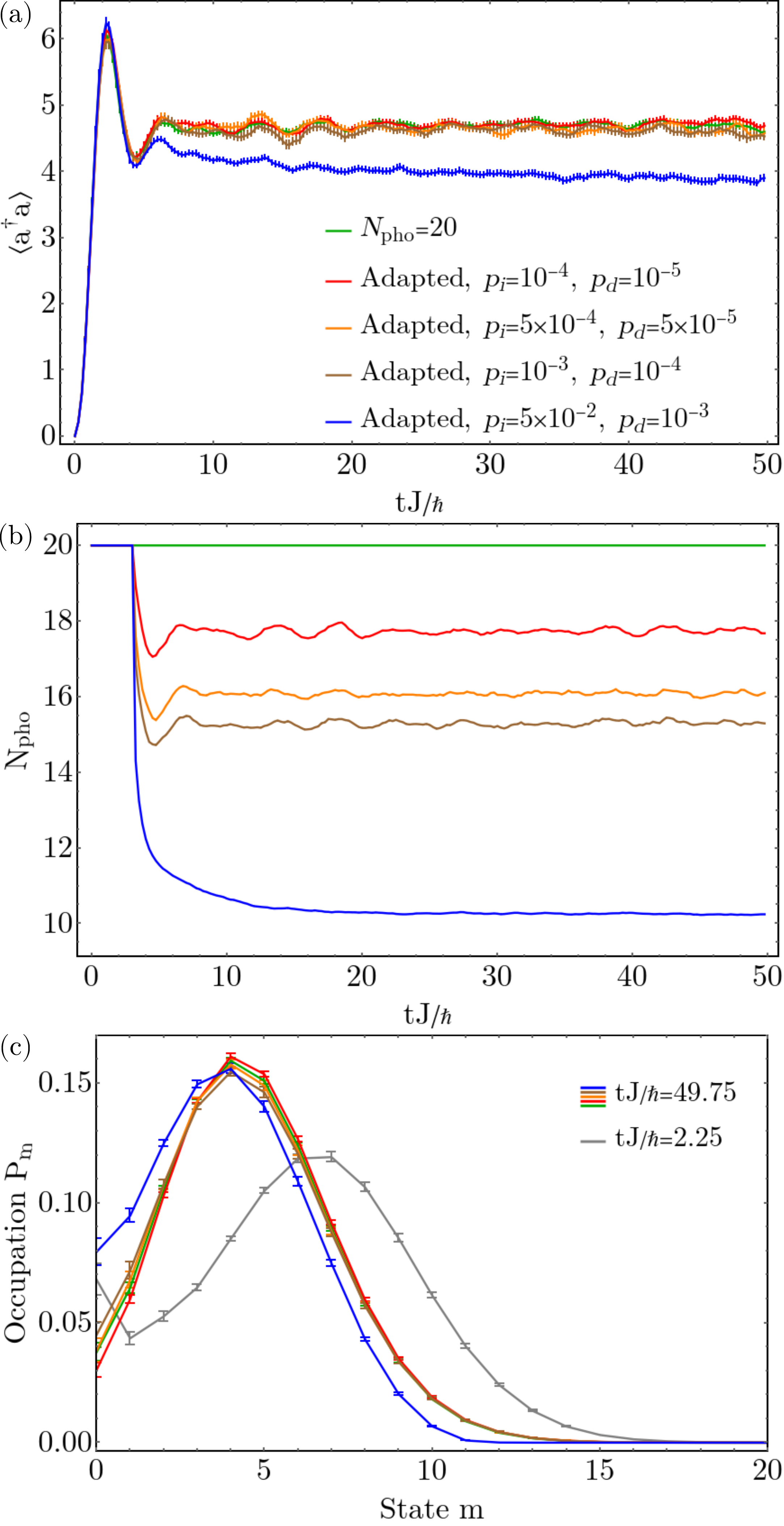}
\caption{
The time evolution of (a) the photon number, $\langle a^\dagger a\rangle$, and (b) $N_\text{pho}$. We compare the results corresponding to an adapted cut-off for different $p_i$ and $p_d$ with the Monte Carlo average of two  different sets of sampled trajectories with a fixed cut-off. 
(c) The photon number distribution, $P_m=\text{tr}\left(\bra{m}\rho\ket{m}\right)$, at $tJ/\hbar=49.75$ for the data presented in (a) and (b), and at $tJ/\hbar=2.25$ with a fixed cut-off. 
The parameters used are the same as Fig.~\ref{fig:adapt}, with $\hbar\Omega\sqrt{N}/J=2.23$.
 }
\label{fig:adapt2}
\end{figure}

In order to implement such an adaptive cut-off, we monitor the evolution of the photon number distribution by measuring the occupation $P_m$ of the photonic Fock states with photon numbers $m$ close to the cut-off value in each time-step. We adapt the cutoff using thresholds for the photonic state occupation. 
To be more precise, the procedure is as following:
At a certain time in the evolution of a single quantum trajectory we have a cut-off $N_\text{pho}(t)$. Depending whether the photon number should increase or decrease we encounter two different cases:
\begin{itemize}
  \item[(i)] The occupation $P_{N_\text{pho}}$ of the photonic Fock states with the largest photon number is smaller than a chosen threshold $p_d$. This signals that the cutoff can be decreased. In order to do this, we find the photonic Fock state $m^*\leq N_\text{pho}$ with the largest photon number whose occupation is above the threshold, i.e. $P_{m^*}\geq p_d$. We change the cut off of the local dimension of the cavity site of the MPS such that the maximal photon number is $N_\text{pho}(t+\mathrm{d}t)=m^*+1$ the for the time step $t+\mathrm{d}t$.
  \item[(ii)] The occupation $P_{N_\text{pho}}$ of the photonic Fock states with the maximum photon number is larger than a second chosen threshold $p_i \geq p_d$, i.e. $P_{N_\text{pho}}\geq p_i$. This signals that the photon number should increase. We increase the local dimension of the cavity site of the MPS at the next time step $N_\text{pho}(t+\mathrm{d}t)=N_\text{pho}(t)+2$.
\end{itemize}

The numerical parameters that control the convergence of the method are now the two threshold values, $p_d$ and $p_i$.

In Fig.~\ref{fig:adapt} and Fig.~\ref{fig:adapt2} we check our procedure of adapting the local dimension for the cavity site by comparing with the results for a fixed converged cut-off. In Fig.~\ref{fig:adapt}(a) and Fig.~\ref{fig:adapt2}(a) we represent the Monte Carlo average of the photon number for two different sets of sampled trajectories with a fixed cut-off and the Monte Carlo average of the photon number with an adapted cut-off for different $p_i$ and $p_d$. We can observe that, except for the case with $p_i=5 \times 10^{-2}$ and $p_d=10^{-3}$, the results for an adapted cut-off agree, within the Monte Carlo averaging error, with the ones for a fixed cut-off. 
The evolution in time of the cut-off can be seen in Fig.~\ref{fig:adapt}(b) and Fig.~\ref{fig:adapt2}(b). We note that at short times, $tJ/\hbar<4$, we always keep the photonic cut-off fixed and relatively large in order to capture the sudden increase in the number of photons in the cavity. But at later times we can observe that in all considered cases the cut-off is approximately $25\%$ smaller than the initial value. This implies a significant speed up of the tMPS method, as the large local dimension of the cavity site being one of the bottlenecks of the method.
As a rough estimate, for the parameters used in Fig.~\ref{fig:adapt} the runtime was with 50\% smaller compared with the case with a fixed cut-off and   for the parameters used in Fig.~\ref{fig:adapt2} with 25\% smaller.
In Fig.~\ref{fig:adapt}(c) and Fig.~\ref{fig:adapt2}(c) we plot the occupation of the photon number states at the final time, $tJ/\hbar=49.75$, to check the agreement of the entire photon number distribution. A very good agreement is found except for the case with $p_i=5 \times 10^{-2}$ and $p_d=10^{-3}$. We also plot in Fig.~\ref{fig:adapt}(c) and Fig.~\ref{fig:adapt2}(c) the photon number distribution at $tJ/\hbar=1.75$, close to the peak in the photon number, to show that at short times many photon number states are occupied.

We note that we also verified the accuracy of this method at the level of single quantum trajectories, not only by analyzing the Monte Carlo average. The improvement brought by this new development is dependent on the physical parameters of the model, but, roughly, has a more important impact when the average photon number is larger, where the difference between the maximum photon number at short times and the steady state value is larger. For example for the parameters in Fig.~\ref{fig:adapt} we manage to lower the local dimension of the cavity site with more than 10 states compared to the fixed cut-off previously used, but for the parameters in Fig.~\ref{fig:adapt2} we have lowered the local dimension with 5 states, due to the lower photon number in the cavity.

\subsection{\label{sec:center}Alternative MPS geometry}

\begin{figure}[!hbtp]
\centering
\includegraphics[width=.5\textwidth]{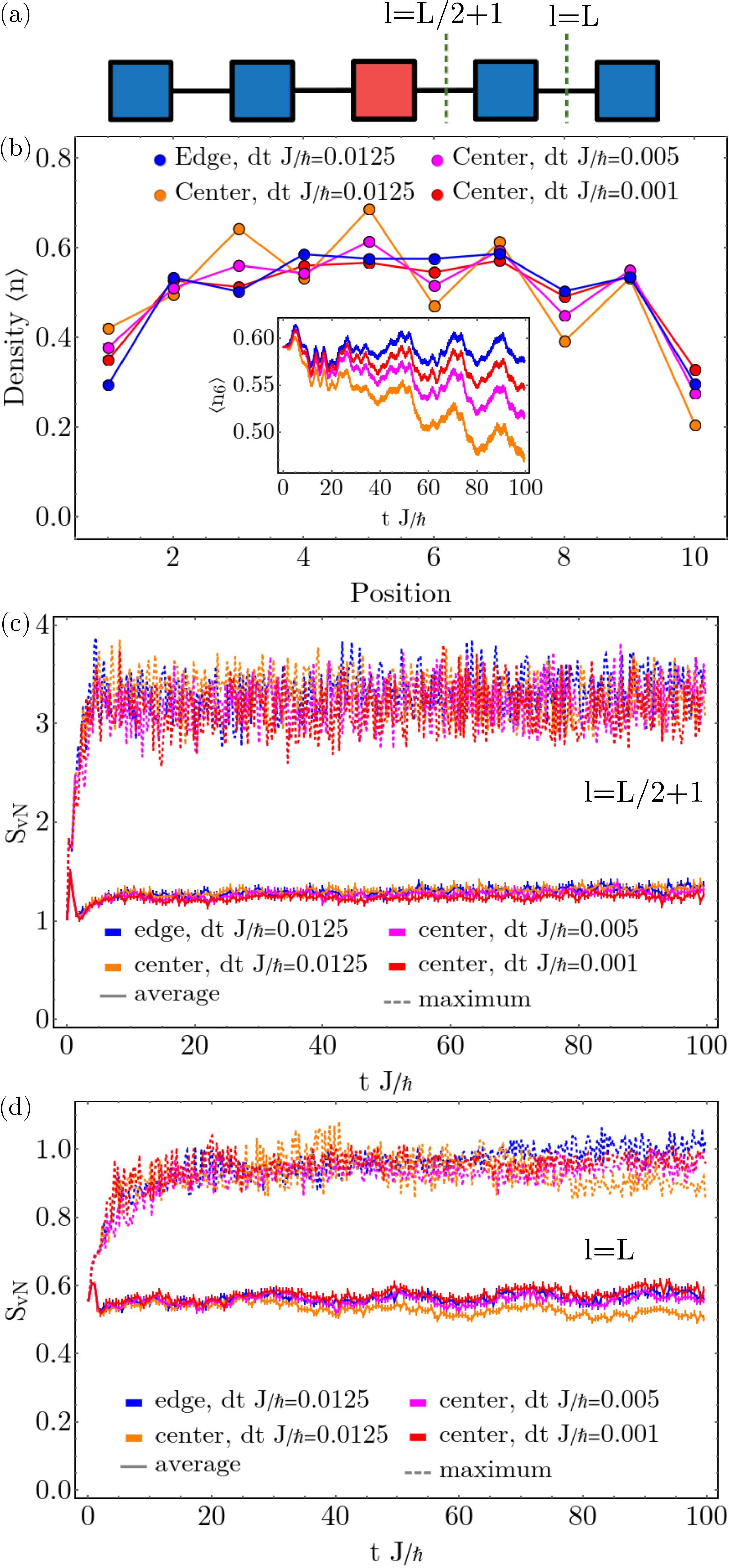}
\caption{
(a) The graphical representation of an alternative MPS structure with the cavity site in the middle of the chain. The bonds for which the von Neumann entropy was computed are marked.
(b) The atomic density profile, $\langle n_i \rangle$, at time $t J/\hbar=99.75$. The inset contains the time evolution of the atomic density $\langle n_6 \rangle$.
(c)-(d) The time evolution of the von Neumann entropy, $S_\text{vN}$ for the Monte Carlo average and the maximum value over different trajectories for (c) $l=L/2+1$ and (d) $l=L$. We compare the two different MPS geometries, with the cavity site at the edge of the atomic chain and the cavity site in the middle of the chain for different time steps $\mathrm{d}t$.
The parameters used are $L=10$, $N=5$, $\hbar\delta/J=2$, $U/J=2$, $\hbar\Omega\sqrt{N}/J=2.46$ and $\hbar\Gamma/J=1$. The Monte-Carlo averages contain at least 500 trajectories.
 }
\label{fig:entcenter}
\end{figure}

The results presented in this work so far have used a MPS geometry in which the cavity site was positioned at the edge of the atomic chain, as depicted in Fig.~\ref{fig:ent1}(a). In this section we compare this approach with a geometry in which the cavity site starts in the middle of the atomic chain, represented in Fig.~\ref{fig:entcenter}(a).
This is motivated by the desire to minimize the occurring entanglement in the system. If the cavity site is positioned in the middle of the chain the average distance between the cavity site and the atomic sites is reduced compared to when the cavity site is at the edge of the chain. This could imply that the entanglement might be reduced and the geometry with the cavity site in the center would be a more efficient representation. However, we show that in the considered cases, the geometry combined with the swap gates is not preferable as described in the following.

The implementation of the new position of the cavity field requires some adaptations compared to the method presented in Sec.~\ref{mps2}. First, the decomposition of the time evolution with the atomic part of the Hamiltonian, given by $e^{-\frac{i\mathrm{d}t}{2\hbar} \left(H_{\text{kin}}+H_{\text{int}}\right)}$, depicted in Fig.~\ref{fig:mps2}, needs to be slighly altered as the atomic sites from the middle of the chain are no longer neighbours in the MPS representation. In order to solve this, one can use the swap gate to bring the two atomic sites next to each other before the application of the time evolution gate. The main difference comes from the application of the operator $e^{-\frac{i\mathrm{d}t}{\hbar} \left(H_{\text{ac}}+H_{\text{c}}-\frac{i}{2}\hbar\Gamma a^\dagger a\right)}$. The decomposition that minimizes the number of swap gates used and leaves the cavity site in the center of the chain is given by

\begin{align}
\label{eq:trott3}
&e^{-\frac{i\mathrm{d}t}{\hbar} \left(H_{\text{ac}}+H_{\text{c}}-\frac{i}{2}\hbar\Gamma a^\dagger a\right)}= \\
&\quad=\prod_{j=1}^{L/2} e^{-\frac{i\mathrm{d}t}{2} \left[-\Omega (a+a^\dagger)(-1)^{j}n_j+\frac{1}{L}\left(\delta-\frac{i}{2}\Gamma\right)  a^\dagger a\right]} \times \nonumber \\
&\quad\quad\prod_{j=L}^{1} e^{-\frac{i\mathrm{d}t}{2} \left[-\Omega (a+a^\dagger)(-1)^{j}n_j+\frac{1}{L}\left(\delta-\frac{i}{2}\Gamma\right)  a^\dagger a\right]} \times \nonumber \\
&\quad\prod_{j=L/2+1}^{L} e^{-\frac{i\mathrm{d}t}{2} \left[-\Omega (a+a^\dagger)(-1)^{j}n_j+\frac{1}{L}\left(\delta-\frac{i}{2}\Gamma\right)  a^\dagger a\right]} +\mathcal{O}(L\mathrm{d}t^2). \nonumber 
\end{align}

We can see that for the MPS structure with the cavity site in the center the Trotter-Suzuki decomposition has a larger error compared to the one used for the geometry with the cavity site at the edge, Eq.~(\ref{eq:trott2}). Thus, we expect that we need a smaller time step, $\mathrm{d}t$, in this case. One can play with the Trotter-Suzuki decomposition at the expense of more swap gates. However, since we show later, that the von Neumann entropy for this geometry is not superior to the previous geometry, we have not gone into this direction. 

In Fig.~\ref{fig:entcenter}(b) we compare the density profile at time $tJ/\hbar=99.75$ obtained with the two geometries. We can observe that in the case with the cavity site at the center the density profile is highly asymmetric using the same time step as for the cavity site at the edge. This signals important errors due to the time step. If we decrease the time step the density profile approaches the one obtained by the implementation with the cavity site at the edge. For a time step of $\mathrm{d}tJ/\hbar=10^{-3}$ the differences are within the Monte-Carlo error for short times, $tJ/\hbar\lesssim 20$, but the deviations at long time are still important, even though for  the case with the cavity site at the edge the time step was more than 10 times larger. We see that obtaining the same accuracy the implementation with the cavity site in the center needs a much smaller time step. 

As one could naively expect that additional numerical effort induced by the smaller required time step to be compensated by the gain in lowering the entanglement, we show in Figs.~\ref{fig:entcenter}(c)-(d) the von Neumann entropy. For the two bipartition considered the von Neumann entropy has very similar values for the two geometries, showing no decrease in the entanglement caused by the shift of the cavity site. We attribute this to the dominating entanglement of the atomic chain and the use of swap gates which brings the cavity site to all different places. Therefore, we also expect that the system size dependence is similar to the one found in Fig.~\ref{fig:ent2} and, in particular, is dominated by the increase of the entanglement of the atoms.

We conclude that we do not expect, as long as we use swap gates, to gain a lot by using different initial geometries and we expect that this further holds for larger system sizes.

\section{Conclusions}

To summarize, we described in detail two methods capable of tackling both short and global range interactions of an interacting  many body system coupled to a single dissipative bosonic mode. We benchmark the methods with the example of a Bose-Hubbard chain coupled to an optical cavity.

The first method is based on the many body adiabatic elimination formalism and applicable for strong dissipation strength. We show how to derive the steady state of the system in the limit of large dissipation strength within the many body adiabatic elimination formalism. The resulting state is a highly mixed state and the reduced density matrix in the atomic sector corresponds to an infinite temperature state. Our results can be evaluated for any system size and we analyzed its physical properties and the thermodynamic limit. In particular, we showed that the obtained steady state has a very different nature compared to the expected mean field state.

As a second algorithm we developed a quasi-exact tMPS method in order to determine the full quantum evolution towards the steady state of the interacting bosonic chain coupled to the cavity. This implementation deals with all the challenges posed by the atom-cavity system: We employ the stochastic unravelling of the master equation to simulate the Lindblad equation, Eqs.~(\ref{eq:Lindblad})-(\ref{eq:Hamiltonian}). The global coupling of the cavity to the atoms is tackled via the dynamical deformation of the MPS structure with swap gates. The efficient simulation of the very large photonic Hilbert space is ensured by its dynamically adapted cut-off. We analyze carefully the convergence of the method for different parameter sets and two different MPS geometries. In particular, we monitored the time dependence of the von Neumann entropy of the quantum trajectory in order to ensure that we properly capture the entanglement between the cavity and the atoms and within the atomic chain.

Both methods open the possibility to treat many body systems coupled to a dissipative bosonic mode beyond the often applied mean field methods. The presented algorithm is easily adapted to fermionic or spin many body systems. Therefore, the methods will have a wide range of application and we expect that many new physical findings will rely on these methods.

\section*{Acknowledgments}

We thank J.-S. Bernier, F. Heidrich-Meisner, {\"O}. Legeza, H.~Ritsch, U.~Schollw\"ock, F. Verstraete, S.~Wolff for stimulating discussions. We acknowledge funding from the Deutsche Forschungsgemeinschaft (DFG, German Research Foundation) in particular under project number 277625399 - TRR 185 (B4) and project number 277146847 - CRC 1238 (C05) and under Germany's Excellence Strategy – Cluster of Excellence Matter and Light for Quantum Computing (ML4Q) EXC 2004/1 – 390534769 and the European Research Council (ERC) under the Horizon 2020 research and innovation programme, grant agreement No.~648166 (Phonton).

\renewcommand{\theequation}{A\arabic{equation}}
\setcounter{equation}{0}

\section*{Appendix A: Many body adiabatic elimination equation of motion}

By writing explicitly the equations of motion, Eq.~(\ref{eq:decfree0}), for the elements of the decoherence free subspace for $\rho^0 = \ket{\alpha(\Delta);n_1,\dotso,n_L}\bra{\alpha(\Delta'=\Delta);n_1',\dotso,n_L'}$ one obtains

\begin{widetext}
\begingroup
\allowdisplaybreaks
\begin{align}
\label{eq:gen_eq_final}
& \pdv{t} \rho^0 = -i(u-u')\rho^0+ J^2e^{-4|\alpha_0|^2}\Bigg\{\\
 & \sum_{i~\text{odd}}\sum_{j~\text{odd}}\Bigg[-\frac{\sqrt{(n_i+1)n_{i+1}}}{\lambda(\Delta-2,u+U(n_i-n_{i+1}+1),\Delta,u')}\times \nonumber \\
& \qquad\qquad\qquad \Bigg(\sqrt{(n_j'+1)n_{j+1}'} \ket{\alpha(\Delta-2);\dotso,n_i+1,n_{i+1}-1,\dotso}\bra{\alpha(\Delta-2);\dotso,n_j'+1,n_{j+1}'-1,\dotso} \nonumber \\
&\qquad\qquad\qquad\qquad +\sqrt{(n_j'+1)n_{j-1}'} \ket{\alpha(\Delta-2);\dotso,n_i+1,n_{i+1}-1,\dotso}\bra{\alpha(\Delta-2);\dotso,n_{j-1}'-1,n_j'+1,\dotso}\Bigg)\nonumber \\
&\qquad\qquad\quad-\frac{\sqrt{(n_i+1)n_{i-1}}}{\lambda(\Delta-2,u+U(n_i-n_{i-1}+1),\Delta,u')}\times \nonumber \\
& \qquad\qquad\qquad \Bigg( \sqrt{(n_j'+1)n_{j+1}'} \ket{\alpha(\Delta-2);\dotso,n_{i-1}-1,n_i+1,\dotso}\bra{\alpha(\Delta-2);\dotso,n_j'+1,n_{j+1}'-1,\dotso} \nonumber \\
&\qquad\qquad\qquad\qquad +\sqrt{(n_j'+1)n_{j-1}'} \ket{\alpha(\Delta-2);\dotso,n_{i-1}-1,n_i+1,\dotso}\bra{\alpha(\Delta-2);\dotso,n_{j-1}'-1,n_j'+1,\dotso}\Bigg)\nonumber \\
&\qquad\qquad\quad-\frac{\sqrt{(n_i'+1)n_{i+1}'}}{\lambda(\Delta,u,\Delta-2,u'+U(n_i'-n_{i+1}'+1))}\times \nonumber \\
& \qquad\qquad\qquad \Bigg( \sqrt{(n_j+1)n_{j+1}} \ket{\alpha(\Delta-2);\dotso,n_j+1,n_{j+1}-1,\dotso}\bra{\alpha(\Delta-2);\dotso,n_i'+1,n_{i+1}'-1,\dotso} \nonumber \\
&\qquad\qquad\qquad\qquad +\sqrt{(n_j+1)n_{j-1}} \ket{\alpha(\Delta-2);\dotso,n_{j+1}-1,n_j+1,\dotso}\bra{\alpha(\Delta-2);\dotso,n_i'+1,n_{i+1}'-1,\dotso}\Bigg)\nonumber \\
&\qquad\qquad\quad-\frac{\sqrt{(n_i'+1)n_{i-1}'}}{\lambda(\Delta,u,\Delta-2,u'+U(n_i'-n_{i-1}'+1))}\times \nonumber \\
& \qquad\qquad\qquad \Bigg( \sqrt{(n_j+1)n_{j+1}} \ket{\alpha(\Delta-2);\dotso,n_j+1,n_{j+1}-1,\dotso}\bra{\alpha(\Delta-2);\dotso,n_{i-1}'-1,n_i'+1,\dotso} \nonumber \\
&\qquad\qquad\qquad\qquad +\sqrt{(n_j+1)n_{j-1}} \ket{\alpha(\Delta-2);\dotso,n_{j+1}-1,n_j+1,\dotso}\bra{\alpha(\Delta-2);\dotso,n_{i-1}'-1,n_i'+1,\dotso}\Bigg)\Bigg]\nonumber \\
&+ \sum_{i~\text{odd}}\sum_{j~\text{even}}\Bigg[\frac{\sqrt{(n_i+1)n_{i+1}}}{\lambda(\Delta-2,u+U(n_i-n_{i+1}+1),\Delta,u')}\times \nonumber \\
& \qquad\qquad\qquad \Bigg(\sqrt{(n_j+1)n_{j+1}}\ket{\alpha(\Delta);\dotso,n_i+1,n_{i+1}-1,\dotso,n_j+1,n_{j+1}-1,\dotso}\bra{\alpha(\Delta);\dotso,n_i',\dotso} \nonumber \\
&\qquad\qquad\qquad\qquad +\sqrt{(n_j+1)n_{j-1}}\ket{\alpha(\Delta);\dotso,n_i+1,n_{i+1}-1,\dotso,n_{j-1}-1,n_j+1,\dotso}\bra{\alpha(\Delta);\dotso,n_i',\dotso}\Bigg)\nonumber \\
&\qquad\qquad\quad+\frac{\sqrt{(n_i+1)n_{i-1}}}{\lambda(\Delta-2,u+U(n_i-n_{i-1}+1),\Delta,u')}\times \nonumber \\
& \qquad\qquad\qquad \Bigg( \sqrt{(n_j+1)n_{j+1}}\ket{\alpha(\Delta);\dotso,n_{i-1}-1,n_i+1,\dotso,n_j+1,n_{j+1}-1,\dotso}\bra{\alpha(\Delta);\dotso,n_i',\dotso} \nonumber \\
&\qquad\qquad\qquad\qquad +\sqrt{(n_j+1)n_{j-1}}\ket{\alpha(\Delta);\dotso,n_{i-1}-1,n_i+1,\dotso,n_{j-1}-1,n_j+1,\dotso}\bra{\alpha(\Delta);\dotso,n_i',\dotso}\Bigg)\nonumber \\
&\qquad\qquad\quad+\frac{\sqrt{(n_i'+1)n_{i+1}'}}{\lambda(\Delta,u,\Delta-2,u'+U(n_i'-n_{i+1}'+1))}\times \nonumber \\
& \qquad\qquad\qquad \Bigg(\sqrt{(n_j'+1)n_{j+1}'}\ket{\alpha(\Delta);\dotso,n_i,\dotso}\bra{\alpha(\Delta);\dotso,n_i'+1,n_{i+1}'-1,\dotso,n_j'+1,n_{j+1}'-1,\dotso} \nonumber \\
&\qquad\qquad\qquad\qquad +\sqrt{(n_j'+1)n_{j-1}'}\ket{\alpha(\Delta);\dotso,n_i,\dotso}\bra{\alpha(\Delta);\dotso,n_i'+1,n_{i+1}'-1,\dotso,n_{j-1}'-1,n_j'+1,\dotso} \Bigg)\nonumber \\
&\qquad\qquad\quad+\frac{\sqrt{(n_i'+1)n_{i-1}'}}{\lambda(\Delta,u,\Delta-2,u'+U(n_i'-n_{i-1}'+1))}\times \nonumber \\
& \qquad\qquad\qquad \Bigg(\sqrt{(n_j'+1)n_{j+1}'}\ket{\alpha(\Delta);\dotso,n_i,\dotso}\bra{\alpha(\Delta);\dotso,n_{i-1}'-1,n_i'+1,\dotso,n_j'+1,n_{j+1}'-1,\dotso} \nonumber \\
&\qquad\qquad\qquad\qquad +\sqrt{(n_j'+1)n_{j-1}'}\ket{\alpha(\Delta);\dotso,n_i,\dotso}\bra{\alpha(\Delta);\dotso,n_{i-1}'-1,n_i'+1,\dotso,n_{j-1}'-1,n_j'+1,\dotso} \Bigg)\Bigg]\nonumber \\
&+ \sum_{i~\text{even}}\sum_{j~\text{odd}}\Bigg[\frac{\sqrt{(n_i+1)n_{i+1}}}{\lambda(\Delta+2,u+U(n_i-n_{i+1}+1),\Delta,u')}\times \nonumber \\
& \qquad\qquad\qquad \Bigg(\sqrt{(n_j+1)n_{j+1}}\ket{\alpha(\Delta);\dotso,n_i+1,n_{i+1}-1,\dotso,n_j+1,n_{j+1}-1,\dotso}\bra{\alpha(\Delta);\dotso,n_i',\dotso} \nonumber \\
&\qquad\qquad\qquad\qquad +\sqrt{(n_j+1)n_{j-1}}\ket{\alpha(\Delta);\dotso,n_i+1,n_{i+1}-1,\dotso,n_{j-1}-1,n_j+1,\dotso}\bra{\alpha(\Delta);\dotso,n_i',\dotso}\Bigg)\nonumber \\
&\qquad\qquad\quad+\frac{\sqrt{(n_i+1)n_{i-1}}}{\lambda(\Delta+2,u+U(n_i-n_{i-1}+1),\Delta,u')}\times \nonumber \\
& \qquad\qquad\qquad \Bigg(\sqrt{(n_j+1)n_{j+1}}\ket{\alpha(\Delta);\dotso,n_{i-1}-1,n_i+1,\dotso,n_j+1,n_{j+1}-1,\dotso}\bra{\alpha(\Delta);\dotso,n_i',\dotso} \nonumber \\
&\qquad\qquad\qquad\qquad +\sqrt{(n_j+1)n_{j-1}}\ket{\alpha(\Delta);\dotso,n_{i-1}-1,n_i+1,\dotso,n_{j-1}-1,n_j+1,\dotso}\bra{\alpha(\Delta);\dotso,n_i',\dotso} \Bigg)\nonumber \\
&\qquad\qquad\quad+\frac{\sqrt{(n_i'+1)n_{i+1}'}}{\lambda(\Delta,u,\Delta+2,u'+U(n_i'-n_{i+1}'+1))}\times \nonumber \\
& \qquad\qquad\qquad \Bigg(\sqrt{(n_j'+1)n_{j+1}'}\ket{\alpha(\Delta);\dotso,n_i,\dotso}\bra{\alpha(\Delta);\dotso,n_i'+1,n_{i+1}'-1,\dotso,n_j'+1,n_{j+1}'-1,\dotso} \nonumber \\
&\qquad\qquad\qquad\qquad+\sqrt{(n_j'+1)n_{j-1}'}\ket{\alpha(\Delta);\dotso,n_i,\dotso}\bra{\alpha(\Delta);\dotso,n_i'+1,n_{i+1}'-1,\dotso,n_{j-1}'-1,n_j'+1,\dotso} \Bigg)\nonumber \\
&\qquad\qquad\quad+\frac{\sqrt{(n_i'+1)n_{i-1}'}}{\lambda(\Delta,u,\Delta+2,u'+U(n_i'-n_{i-1}'+1))}\times \nonumber \\
& \qquad\qquad\qquad \Bigg(\sqrt{(n_j'+1)n_{j+1}'}\ket{\alpha(\Delta);\dotso,n_i,\dotso}\bra{\alpha(\Delta);\dotso,n_{i-1}'-1,n_i'+1,\dotso,n_j'+1,n_{j+1}'-1,\dotso} \nonumber \\
&\qquad\qquad\qquad\qquad+\sqrt{(n_j'+1)n_{j-1}'}\ket{\alpha(\Delta);\dotso,n_i,\dotso}\bra{\alpha(\Delta);\dotso,n_{i-1}'-1,n_i'+1,\dotso,n_{j-1}'-1,n_j'+1,\dotso} \Bigg)\Bigg]\nonumber \\
&+ \sum_{i~\text{even}}\sum_{j~\text{even}}\Bigg[-\frac{\sqrt{(n_i+1)n_{i+1}}}{\lambda(\Delta+2,u+U(n_i-n_{i+1}+1),\Delta,u')}\times \nonumber \\
& \qquad\qquad\qquad \Bigg(\sqrt{(n_j'+1)n_{j+1}'} \ket{\alpha(\Delta+2);\dotso,n_i+1,n_{i+1}-1,\dotso}\bra{\alpha(\Delta+2);\dotso,n_j'+1,n_{j+1}'-1,\dotso} \nonumber \\
&\qquad\qquad\qquad\qquad+\sqrt{(n_j'+1)n_{j-1}'} \ket{\alpha(\Delta+2);\dotso,n_i+1,n_{i+1}-1,\dotso}\bra{\alpha(\Delta+2);\dotso,n_{j-1}'-1,n_j'+1,\dotso}\Bigg)\nonumber \\
&\qquad\qquad\quad-\frac{\sqrt{(n_i+1)n_{i-1}}}{\lambda(\Delta+2,u+U(n_i-n_{i-1}+1),\Delta,u')}\times \nonumber \\
& \qquad\qquad\qquad \Bigg(\sqrt{(n_j'+1)n_{j+1}'} \ket{\alpha(\Delta+2);\dotso,n_{i-1}-1,n_i+1,\dotso}\bra{\alpha(\Delta+2);\dotso,n_j'+1,n_{j+1}'-1,\dotso} \nonumber \\
&\qquad\qquad\qquad\qquad+\sqrt{(n_j'+1)n_{j-1}'} \ket{\alpha(\Delta+2);\dotso,n_{i-1}-1,n_i+1,\dotso}\bra{\alpha(\Delta+2);\dotso,n_{j-1}'-1,n_j'+1,\dotso} \Bigg)\nonumber \\
&\qquad\qquad\quad-\frac{\sqrt{(n_i'+1)n_{i+1}'}}{\lambda(\Delta,u,\Delta+2,u'+U(n_i'-n_{i+1}'+1))}\times \nonumber \\
& \qquad\qquad\qquad \Bigg(\sqrt{(n_j+1)n_{j+1}} \ket{\alpha(\Delta+2);\dotso,n_j+1,n_{j+1}-1,\dotso}\bra{\alpha(\Delta+2);\dotso,n_i'+1,n_{i+1}'-1,\dotso} \nonumber \\
&\qquad\qquad\qquad\qquad +\sqrt{(n_j+1)n_{j-1}} \ket{\alpha(\Delta+2);\dotso,n_{j-1}-1,n_j+1,\dotso}\bra{\alpha(\Delta+2);\dotso,n_i'+1,n_{i+1}'-1,\dotso} \Bigg)\nonumber \\
&\qquad\qquad\quad-\frac{\sqrt{(n_i'+1)n_{i-1}'}}{\lambda(\Delta,u,\Delta+2,u'+U(n_i'-n_{i-1}'+1))}\times \nonumber \\
& \qquad\qquad\qquad \Bigg(\sqrt{(n_j+1)n_{j+1}} \ket{\alpha(\Delta+2);\dotso,n_j+1,n_{j+1}-1,\dotso}\bra{\alpha(\Delta+2);\dotso,n_{i-1}'-1,n_i'+1,\dotso} \nonumber \\
&\qquad\qquad\qquad\qquad +\sqrt{(n_j+1)n_{j-1}} \ket{\alpha(\Delta+2);\dotso,n_{j-1}-1,n_j+1,\dotso}\bra{\alpha(\Delta+2);\dotso,n_{i-1}'-1,n_i'+1,\dotso} \Bigg)\Bigg]\Bigg\}.\nonumber 
\end{align}
\endgroup

\end{widetext}
In order to identify the steady states we need to solve $\pdv{t} \rho^0=0$.  If we look at the coefficients of the diagonal terms of the decoherence free subspace, $\ket{\alpha(\Delta);n_1,\dotso,n_L}\bra{\alpha(\Delta);n_1,\dotso,n_L}$, in Eq.~(\ref{eq:gen_eq_final}), we observe that their sum is zero, thus the mixed state given by
\begin{align}
\label{eq:ss_gen2}
\rho_{\text{mix}} &=\frac{1}{\mathcal{N}}\sum_{\{n_j\}} \ket{\alpha(\Delta);n_1,\dotso,n_L}\bra{\alpha(\Delta);n_1,\dotso,n_L}
\end{align}
is a steady state of the system, where we sum over all possible density configurations $\{n_j\}$ and $\mathcal{N}$ is the number of these configurations, $\mathcal{N}=\begin{pmatrix} L+N-1 \\ N \end{pmatrix}$.

\end{document}